\documentclass[letterpaper,12pt,leqno]{article}
\usepackage{paper}
\usepackage[section]{placeins}
\usepackage{booktabs}

\hypersetup{pdftitle={Paper Example}}

\usepackage{tkz-graph}
\usetikzlibrary{positioning}
\usetikzlibrary{backgrounds}
\usetikzlibrary{arrows.meta}
\tikzstyle{observed}=[circle, inner sep=0mm, outer sep=0mm, minimum size=2mm, draw=black, fill=black]
\tikzstyle{unobserved}=[circle, inner sep=0mm, outer sep=0mm, minimum size=2mm, draw=black, fill=white]
\tikzstyle{phantom}=[inner sep=0mm, outer sep=0mm, minimum size=2mm, draw=black, fill=black]
\tikzstyle{error}=[circle, inner sep=0mm, outer sep=0mm, minimum size=2mm, draw=black, fill=white]
\tikzstyle{notouch}=[shorten <=5pt, shorten >= 5pt, -{Latex[length=2mm, width=1.5mm]}]
 
\usepackage{mathtools}
\usepackage{graphicx}
\graphicspath{{mainfigures/}}

\usepackage[colorinlistoftodos]{todonotes}

\newtheorem{example}{Example}

\newcommand{\indep}{\perp\!\!\!\perp}

\newcommand{\pr}{{\rm pr}}

\newcommand{\bib}{refs.bib}

\begin{document}

\title{\bf Estimating Causal Effects from Data Generated by Stochastic Algorithms}
\author{Susan Athey, Guido 
Imbens, and Zoe Ji\thanks{
Susan Athey, Graduate School of Business, Stanford University;
Guido Imbens, Graduate School of Business and Department of Economics, Stanford University;
Zoe Ji, Institute of Computational and Mathematical Engineering, Stanford University.
This work was partially supported by the Office of Naval Research under grant N00014-17-1-2131 and a gift from Amazon.
}}
\date{July 2026}   
\begin{titlepage}
\maketitle
\vspace*{-1.1cm}
\begin{abstract} 

Recommendation systems and chatbots present content to users, typically using stochastic algorithms that select the content based on user characteristics or context. Examples of content include chat responses, videos, or items available for purchase. Scientists and application developers are often interested in whether characteristics of content increase outcomes such as user engagement. Estimates of such causal effects may guide
content providers to generate content that emphasize desirable features.
However, in settings with a large content library or where content is generated uniquely for a given user, it can be difficult to use observational data to learn the causal effect of content features, because the content a user sees is tailored to that user, and because content varies in many dimensions. 
The basic observation that motivates this paper is that because modern AI systems select and generate content stochastically, routine interactions between users and these systems continuously produce micro experiments, and logging two additional elements of each interaction turns those micro experiments into  data that identify causal effects of content features.
This paper proposes a new method for identifying and estimating the impact of content features using observational data, when the algorithm that determines user exposure incorporates some randomization, and when two additional data elements are logged for each user: $(i)$ the identity of at least one item that could have been exposed to the user, but was not (the unexposed item); $(ii)$ an estimate of the ratio of the probability that the unexposed item would have been shown to the probability that the exposed item was shown.

\end{abstract}
\end{titlepage}


\clearpage

\section{Introduction}

Generative artificial intelligence (AI) algorithms are used to create content such as chatbot responses or video in response to user queries and context. In digital applications such as social media and e-commerce, AI-powered recommendation systems operate behind the scenes to select content based on user queries and/or context. In both cases, the content exposed to users is personalized.  In addition, the exposed content is not a deterministic function of the user query and context, that is, the algorithm is stochastic. In the case of generative AI, the randomness may derive from the stochastic nature of the underlying large language model or diffusion model; and for most online digital applications, additional sources of randomness arise from ongoing experimentation and bandit algorithms as well as randomness in supporting algorithms, server updates and other operational processes. In this paper, we develop new methods for estimating the causal effects of content features using observational data generated by stochastic algorithms. Examples of content features may include the style or tone of chatbot responses, characteristics of user-generated content in social media such as image features, or characteristics of product listings such as the number of bedrooms for lodging platforms. Content varies in many such dimensions at once. For expositional purposes, we focus on the causal effect of a single target feature that takes on two values, low and high, while the other dimensions of the content are not held fixed and instead vary naturally with the target feature, as they do in the content the algorithm produces.

From a scientific perspective, an estimate of the causal effect of a content feature may be used to test hypotheses about consumer behavior or consumer psychology, for example hypotheses about the impact of emotional language in textual descriptions or responses or hypotheses about a consumer's willingness-to-pay for quality. Other hypotheses may be relevant to the study of educational techniques; for example, for an educational chatbot, content features might include distinct pedagogical approaches. 

Estimates of the causal effects of content features are also useful for application developers.  Although our methods apply to any measurable features of content, they may be particularly useful when the features are selected to be interpretable and actionable. For example, if an analyst identifies a feature that can be succinctly communicated in an LLM system prompt, such as friendliness, then a finding that user engagement would be increased if the feature increased can be translated into an intervention that modifies the system prompt of the LLM.  If an analyst for an online service estimates a positive effect on engagement when the service provider smiles in the profile picture (as in \cite{athey2022smiles}), the marketplace can implement an intervention where service providers are nudged to smile in the photo. A finding that a feature of a product listing has a positive effect may enable an e-commerce platform or marketplace to recruit more product listings with the desired features.  

In practice, several challenges arise in estimating the causal effects of content features. The first is a direct consequence of personalization: different users are shown different content. In the case of generative AI or sophisticated digital applications, each user context and query is effectively unique. Personalization creates well-known challenges for causal inference. For example, a user with a history of asking about homework questions may receive a different response than a user with a history of asking about relationships. If the content feature of interest is the inclusion of factual information in the response, and users doing homework have higher engagement, observational data will reveal a positive correlation between factual information and engagement. Yet, increasing the factual information for relationship queries may not improve user engagement. In this example, the type of query is a confounder from the perspective of estimating the causal effect of factual content. This example also surfaces further complications: factual content has heterogeneous effects across user contexts and queries, but further, factual information may be beneficial when bundled with certain other features of a chatbot response, but annoying when bundled with alternative features, and these other features covary systematically with factual information in observational data. An empathetic response might become annoying to users when combined with facts.

There are a variety of approaches in the existing literature for estimating causal effects of features  (including the literature on off-policy evaluation based on unconfoundedness approaches \cite{schnabel2016unbiased, joachims2017unbiased, uehara2022review} and approaches based on the negative sampling approach \cite{he2016vbpr}), but each of them faces challenges in settings with high-dimensional or unique user contexts and extremely large and heterogeneous sets of potential content. One approach is to run a randomized experiment varying the feature of interest. This approach faces challenges because content varies in many dimensions, and the effect of a given feature may be different when it coincides with other dimensions of content. Although an experiment could attempt to vary many features independently to learn the causal effect of each (a factorial design), such an experiment would require a large sample to accommodate interactions among many features. Further, some combinations of features might not make sense and could thus be expensive from a user experience perspective (a problem that motivates the use of contextual bandits, discussed in more detail below).  A second is to make use of observational data, thus expanding sample size, while attempting to adjust for confounding when estimating causal effects. This type of approach faces challenges due to the difficulty of fully adjusting for all aspects of user context that matter for confounding. Further, personalized algorithms expose users to content that varies in many dimensions beyond the feature of interest, dimensions that are correlated with the feature of interest and thus also confound treatment effect estimation. 
 
In this paper, we propose a new method for estimating the impact of content features using observational data generated from user interactions with personalized stochastic algorithms. Our approach requires that additional data elements be logged for each user. In the base case, two data elements are logged: $(i)$ the identity of at least one item of content that could have been exposed to the user, but was not (the unexposed content); $(ii)$ an estimate of the ratio of the probability that the unexposed content would have been shown, relative to the probability of the exposed content being shown. Notably, the outcome for the unexposed content is not observed, since it was never exposed to a user. We refer to datasets that log the features of the content exposed to the user, the user's outcome, alongside $(i)$ and $(ii)$ as data with ``logged counterfactual exposures'' (LCE). We show that causal effects of features are identified given data with LCE, even in the presence of unobserved confounders that affect both user preferences and the identity of the considered set of content (exposed and unexposed). Our estimator differs from existing approaches in terms of what data is used ($(i)$ and $(ii)$ are not typically incorporated), and our estimator is novel in the way that the observational data is weighted and averaged to construct an average causal effect. The definition and analysis of LCE data is the first contribution of this paper. 

In settings where $(ii)$ is not available (relative probabilities of items are not logged), another type of LCE data incorporates a larger set of counterfactual exposures.  For each user context and query, a set of counterfactual items of content is sampled from the stochastic algorithm, and features of interest are recorded for the counterfactual exposures. To apply our approach in this setting, it is not necessary to estimate a propensity model that maps user characteristics to content assignment or to feature values, nor is it necessary to record or adjust for all features of content that are related to both user context and user outcomes.

In our second contribution we show how LCE data can be used to estimate well-defined causal effects, even in the absence of logged data about user characteristics and other content features. Making this claim precise requires care about what ``the causal effect of a feature'' means in this setting. In our framework, the treatments are the items of content themselves: potential outcomes are defined by the content a user is exposed to, not by its measured features, because items of content differ in many dimensions, only some of which are recorded. While content can be classified according to an individual feature, the content as a whole (including its measured and unmeasured features) determines outcomes. For a given (exposed, non-exposed) pair of content and a given user, the causal effect of exposing the user to the item of content with a high value rather than the low value of the target feature is a well-defined causal effect, but it is specific to that particular (exposed, non-exposed) pair: the two items of content differ in the target feature, but the target feature tends to covary with other measured or unmeasured characteristics of the content. 

Our estimands, which we call conditional treatment-averaged causal effects (C-TACEs), average these user-specific contrasts in two directions: over users, as is standard in the causal inference literature, and over the pairs of items of content that the stochastic algorithm generated for those units. The second averaging is novel, and the interpretation is subtle. The C-TACE is the average effect of receiving high-feature content as the deployed algorithm actually produces it, inclusive of the other content dimensions that are correlated with the feature in that algorithm's output, rather than the effect of an isolated change in the feature holding all other dimensions fixed. For a platform deciding whether to steer its algorithm to increase the value of the feature, the former is policy-relevant; we discuss the relation between the two, and the conditions under which they coincide, in Section \ref{sec:linear-example}.

To further elucidate the interpretation of the causal effect captured by C-TACE, we introduce the notion of {\it micro experiments}, experiments of size one, where a single unit is exposed to one level of a target feature but could have been exposed to one or more known alternative levels of a target feature.
These micro experiments are {\it doubly heterogeneous}. First of all, they are heterogeneous in the units participating in the experiments. This type of heterogeneity is widely recognized in the causal inference field and has led to the literature on estimating heterogeneous causal effects \citep{wager2018estimation}.
Second, these micro experiments are heterogeneous in the content the units are potentially exposed to. Content varies in many dimensions that are correlated with one another, with the content feature of interest, and with the user's outcomes. Our approach is compatible with a setting where each item of content is only exposed to one user in one context.

In our third contribution we discuss the prevalence of LCE data, where a particularly relevant and novel application of the methods is to data from LLMs or other generative models. A typical chat application or product feature collects context and a user's query and passes those to an API for a cloud-hosted LLM, which then passes back a response. Most cloud-hosted open-weight LLMs and some frontier closed-weight LLMs ({\it e.g.} models from OpenAI and Google) expose the option to return the log-probabilities of the response tokens alongside the response. Doing this for the answer exposed to the user would yield logged bandit feedback data, albeit with very small probabilities for a given piece of content. In order to generate LCE data for a sample of user exposures, the analyst could make two API calls instead of one, recording for both the exposed content and the unexposed content the log probability of the response and normalizing by the sum of the two. This doubles the API usage cost relative to logged bandit feedback data.\footnote{In general, the second API call could take place offline using batch processing for a selected sample of queries so as to reduce cost and avoid slowing down the user response in the online interaction.} If log-probabilities are not exposed, LCE data can be created with additional sampling. For each user context and query, the analyst can sample a small set of additional responses and use the data to estimate the probability that the algorithm exposed the user to a response with the feature value that was in fact exposed to the user (note that this is distinct from the probability that the specific response was shown). Unlike common practice for recommendation systems, LLM providers have maintained access to many (but not all) historical versions of LLMs for a period of time, making it more practical to resample from the same stochastic algorithm without concern that the algorithm changes between samples. This approach does require the analyst to save the context and query that was initially passed to the LLM, but it does not require the analyst to build a full model of assignment using that context.

Now consider how LCE data may arise in the context of recommendation systems. One potential source is data from randomized experiments. Although it is relatively rare for academic researchers to access data from large-scale randomized experiments, thousands of such experiments are run per year inside large technology firms \citep{kohavi2013online, gupta2019top}.  There are several types of randomized experiments relevant for estimating effects of features. In the first, we randomize the content exposed drawing from a fixed content library, with the goal of collecting data that can be used to learn a personalized content assignment policy. \cite{bottou2013counterfactual} imposed randomization of position among the top-ranked advertisements at the Bing search engine in order to collect data to estimate causal effects of position. \cite{li2010contextual} analyzed a contextual bandit that was used to select which news story to display from a set of a few news stories on the home page of Yahoo! News. The literature focused on analyzing these experiments introduced the term ``logged bandit feedback'' to describe the case where the data includes the assignment probability (the propensity score) for the item that was exposed to the user. Universal logging of the propensity score is not standard practice in technology firms because the computation is expensive when the catalog of content is large; often stochastic algorithms are designed to sample an item from a distribution rather than calculate probabilities. 

For the case of generative AI, the catalog of content is effectively infinite. In contrast to logged bandit data, LCE data would require data about the features of content that was not shown, as well as the relative probability of the exposed versus unexposed content. Non-exposed content is often logged when recommendation systems proceed in multiple stages, where an initial stage uses inexpensive computation to narrow down a candidate set of content, and then more expensive computation is used to select the best choice from the candidate set. Logging the relative probability of non-exposed content is by no means universal, but it would be known in a given experiment. Introducing such logging is relatively inexpensive, because only two pieces of content need to be analyzed in-depth. In addition, if the logging was not done in the context of a recommendation system, data from (exposed, unexposed) pairs can be used to estimate, as a function of user context and query, how likely the user is to be exposed to content with high levels of the feature of interest.

One algorithm may introduce a new feature to the recommendation system or place more emphasis on a given feature, and so the experiment indirectly supports an estimate of the effect of the feature. Formally, the algorithm assignment is an instrumental variable that affects assignment to a high level of the feature but does not directly affect outcomes \citep{hartford2017deep}.  In the case of an LLM, a change to a system prompt to, e.g., promote friendliness is analogous to introducing a new feature in a recommendation system; it increases the average value of the feature overall, but does not increase friendliness in every query. Related types of interventions include experiments that randomize users to an alternative user experience that elevates certain information, or experiments that randomize a nudge provided to suppliers in a marketplace.

Experiments of these kinds can be a valuable source of LCE data. 
However, data from well-conducted experiments are rarely available to scientific researchers; even within technology companies they may be limited in scope, and they often include only limited variation in the interventions of interest.
The observation that motivates this paper is that no purpose-built experiment is required: because modern AI systems select and generate content stochastically, routine interactions between users and these systems continuously produce micro experiments, and logging two additional elements of each interaction turns those micro experiments into LCE data that identify causal effects of content features.

\section{Related Literature}

There is a large literature on estimating the effects of the design of recommender systems and other content-selection algorithms. Part of this literature compares algorithms directly in randomized experiments. Another literature draws inferences that are not directly validated by an A/B test, and it is that part that is closest to our setting. The literatures reviewed below share a common structure: each works with the action that was chosen and some version of its propensity, whether a marginal probability, a generalized propensity, or an embedding-level probability, and then either assumes unconfoundedness given the observed context, restores unconfoundedness by design through deliberate randomization, or accepts partial identification under a sensitivity model. In contrast, our approach relaxes strong assumptions while enabling identification through additional logged data.   

One example  is the literature on combining multiple experiments comparing algorithms to learn about an effect that was not the focus of the experiments. \cite{hartford2017deep} combined data from a number of experiments that evaluate ranking algorithms. 
In their analyses they focus on estimating the effect of moving an item from one position to another position in the display. They use the indicator for the experiment as an instrument for the position.

A second literature studies off-policy evaluation and learning from logged bandit feedback. For each interaction the data record the context, the action chosen by a stochastic logging policy, the propensity of that action under the policy, and the realized reward; the reward of any action other than the one chosen is missing. \citet{swaminathan2015counterfactual, swaminathan2015batch} introduce counterfactual risk minimization, which minimizes a variance-regularized inverse-propensity-weighted estimate of the risk, and \citet{swaminathan2015self} correct a propensity-overfitting pathology of inverse-propensity weighting through self-normalization. \citet{dudik2011doubly} give a doubly robust estimator that is consistent if either the propensity model or the reward model is correct, and \citet{wang2017optimal} characterize the minimax evaluation error and switch between the doubly robust and the direct estimate in the region where importance weights are large. A review of these methods and their assumptions is given by \citet{uehara2022review}. All of them log the marginal propensity of the single chosen action and identify effects under unconfoundedness given the logged context; none logs the identity of a specific alternative or the relative probability of a pair.

The action-level propensity becomes uninformative when the action space is large, because the probability of any single action is small and the importance weights are unstable. \citet{saito2022off, saito2023off} reweight instead by the propensity of an action's embedding, which identifies the target under the assumption that the reward depends on the action only through the embedding. This is the closest antecedent to our setting, in which the actions are items of content that are effectively unique. The methods differ in what restores identification: embedding-level reweighting requires that measured features mediate the reward, whereas our identification uses the relative propensity of a specific logged pair and does not require that assumption.

The randomization that these off-policy methods exploit arises from several sources. \citet{bottou2013counterfactual} inject controlled randomization into the ad-placement scores of a search engine so that the resulting logs support counterfactual estimates, with propensities known by construction, and argue that the offline evaluation of an interactive system requires causal rather than predictive reasoning. \citet{li2010contextual} develop contextual-bandit algorithms for news recommendation, and \citet{li2011unbiased} show that serving a uniformly random slice of traffic permits unbiased offline evaluation of any policy by rejection sampling on the logs. \citet{narita2021algorithm} observe that any known algorithmic decision rule with a stochastic or threshold component is itself an experiment, so that its propensities are recoverable from the rule, while \citet{langford2008exploration} and \citet{narita2019efficient} make use of randomization already present in existing logs. Public datasets constructed on the same principle include those of \citet{lefortier2016large}, \citet{gilotte2018offline}, and \citet{saito2020open}. Our setting relies on the same type of randomization, but requires only the relative probability of a considered pair, which a stochastic algorithm can report even when the marginal probability of a single item is not logged.

In recommender systems, a related literature uses records of what was and was not shown, but to a different end: constructing training labels for preference models rather than identifying causal effects. Implicit-feedback data record interactions only for exposed items, so that a non-interaction confounds a disliked item with an unseen one, and the exposure-bias literature corrects this missing-not-at-random structure through propensity weighting and exposure modeling \citep{schnabel2016unbiased, joachims2017unbiased, saito2021counterfactual}. The negative-sampling literature constructs negative examples for training, either by drawing from the catalog, which conflates an unexposed item with a disliked one, or by using impressions, items that were shown but not clicked and that the serving system logs \citep{he2016vbpr, ding2018improved, ding2019reinforced, ma2024negative}. That platforms already log impressions at scale supports the feasibility of our logging requirement, but the roles differ: in negative sampling the unexposed item is a label for supervised learning, whereas in our data structure it is a counterfactual exposure whose relative propensity is the identification device. Nothing in this literature logs or uses the exposure-probability ratio of a specific pair, and none of it targets causal effects of content features.

A semiparametric literature develops methods for continuous actions, where per-action propensities are undefined. \citet{chernozhukov2019semi} and \citet{syrgkanis2023post} build doubly robust, Neyman-orthogonal estimators on a generalized propensity, the conditional density of the action, using flexible machine learning for the nuisance functions, and \citet{kallus2018balanced} choose evaluation weights to optimize balance directly rather than inverting an estimated propensity. The standard route out of a finite action space is thus to replace propensity mass functions with conditional densities; our route avoids estimating any density over the content space by logging a relative propensity for a specific pair.

Another strand confronts unobserved confounding directly. \citet{bennett2019policy} evaluate policies when only proxies for a latent confounder are observed, and \citet{kallus2021minimax, kallus2022doubly} posit a sensitivity model that bounds the deviation between nominal and true propensities and learn the policy that maximizes worst-case value. This line accepts that the value of a policy is only partially identified and optimizes over the resulting identified set. Our claim is different in kind: the two additional logged elements restore point identification of content-feature effects even when unobserved confounders drive both user preferences and which pair of items the algorithm considered. The comparison must be drawn carefully, because the estimands differ, a local contrast between the features of a considered pair rather than the value of a policy, and because our identification relies on the algorithm's internal randomization being recorded, which is not available in the observational settings these papers study.

A final distinct literature studies adaptively collected data, in which the logging policy changes over time, so that assignment probabilities are history-dependent and observations are dependent. \citet{hadad2021confidence} and \citet{zhan2021off} recover valid inference by adaptively reweighting augmented inverse-propensity scores to restore a central limit theorem; \citet{zhan2024policy} extend the argument to policy learning under a decaying exploration floor; \citet{zhou2023offline} give efficient multi-action policy learning in the independent benchmark; and \citet{dimakopoulou2019balanced, dimakopoulou2018estimation} build balancing into the bandit's own estimation. The concern in this literature is the dependence induced by adaptivity, which is orthogonal to our problem: we take the relative propensity of the pair as recorded, and the identification argument does not confront history-dependence.

A distinct literature studies causal effects when a nominal treatment has multiple unobserved versions, or is an aggregate of finer components. \citet{hernan2011compound} and \citet{vanderweele2013causal} formalize a compound treatment whose realized version varies, and show that when the versions are not modeled, the identified quantity is a version-mixed effect: a contrast that averages the version-specific effects over the distribution of versions as they occur, under the assumption that version assignment is unconfounded given the observed covariates. \citet{tsao2026lost} sharpen the ambiguity for aggregate treatments, showing that the effect of intervening on an aggregate depends on how the intervention is instantiated across its components, formalized as a distribution over component values consistent with the aggregate, and that standard instrumental-variable estimators recover a component-instantiation-specific effect only under restrictive conditions. Our estimand shares the structure of these quantities. A content feature does not determine the item shown, so the item plays the role of an unobserved version, or of the components underlying an aggregate feature, and the C-TACE averages the item-level contrasts over the versions the algorithm produces rather than over a prespecified or hypothetical instantiation distribution. The approaches part on identification. The version-mixed and aggregate effects are identified by adjusting for the covariates that confound version assignment, or by an instrument satisfying an exclusion restriction, both of which constrain the observed data; our identification uses the recorded within-pair randomization, and therefore holds even when the variables driving which pair of items the algorithm considered are never observed. The instantiation distribution is not a modeling choice to be defended but the deployed algorithm's own generation process, which is also the distribution a platform intervening on the feature would act through.

The 
fractional factorial design literature, like our setting, involves averaging over different treatments, \cite{dasgupta2015causal, de2022improving, xuan2025evaluating}. In the factorial design setting there is typically only a limited number of treatments.

The multi-armed bandit literature \citep{lattimore2020bandit} has explicit randomization of the type we exploit here, but typically focuses on a modest number of treatments.

Neither the factorial-design literature nor the bandit literature directly corresponds to the setting we study, but both show that it is non-trivial to set up the formal problem of policy evaluation and policy estimation, and to distinguish the data-generating process from the estimation target. See also
\cite{athey2026targeting, agrawal2026economics}.

\section{Set Up}\label{sec:setup}

We consider a setting where each unit, indexed by $i$, can be exposed to one of a potentially very large set of treatments, denoted by ${\cal A}$. 
Associated with each treatment $a\in{\cal A}$  is a {\it potential outcome} $Y_i(a)\in\mathbb{R}$ if unit $i$ is exposed to treatment $a$ \citep{ imbens2015causal}. 
However, assignment to the treatments in this set is not the same for all units.
Unit $i$ is  randomly assigned to a treatment that is an element of a {\it unit-specific treatment set} ${\cal A}_i$ that is a subset of the full set of treatments, ${\cal A}_i\subset {\cal A}.$
This  { unit-specific treatment set} ${\cal A}_i$ for unit $i$ may vary systematically by unit, as captured by its indexing by $i$. In particular the set ${\cal A}_i$ may be related to the potential outcomes $Y_i(a)$. For example, if for some treatment $a$ the value $Y_i(a)$ is high, this treatment may be more likely to be in the treatment set ${\cal A}_i$ than a treatment value $a'$ such that $Y_i(a')$ is low. 

For expository reasons we focus initially  on the case where  the cardinality of each unit-specific treatment set ${\cal A}_i$ is two, ${\cal A}_i=\{A_i(0),A_i(1)\},$ but our results extend naturally to cases where the cardinality is higher, and we discuss such cases in Section \ref{section:multiple_treatments}.
Each unit also has a set of characteristics $X_i$ not affected by the treatment, but possibly related to the potential outcomes and to the treatment set. 

A key assumption is that each unit is randomly assigned to one of the elements on the unit-specific treatment set. Let  $W_i\in\{0,1\}$ denote the indicator for the assignment. 

\begin{assumption}\label{ass:randomization}{\sc (Randomized Assignment)}
\begin{equation}
    \pr\Bigl(W_i=1\Bigl| A_i(0),A_i(1),Y_i(A_i(0)),Y_i(A_i(1)),X_i\Bigr)=1/2.
\end{equation}    
\end{assumption}
In Section \ref{section:nonequal} we generalize this to the case where units are assigned to one of the treatments in their treatment set ${\cal A}_i$ with known probabilities, possibly different from $1/2.$

The realized treatment is $A_i=A_i(W_i)$, with the non-realized treatment $A^N_i=A_i(1-W_i)$, and the realized and non-realized outcomes are $Y_i=Y_i(A_i(W_i))$ and $Y_i^N=Y_i(A_i(1-W_i))$. 
 In much of the causal inference literature, {\it e.g.,} \cite{imbens2015causal}, the non-realized potential outcomes $Y_i(A_i(1-W_i))$ are referred to as ``missing'' potential outcomes to contrast them with the observed values $Y_i=Y_i(A_i(W_i))$. Here it is crucial that some of these non-realized potential outcomes are in fact observed, so we use the ``non-realized'' label instead of the ``missing'' label.

Let $f:{\cal A}\mapsto \mathbb{R}^d$ be a mapping from the space of treatments into a $d$-dimensional real vector of
 {\it treatment features}, and let $V_i=f(A_i)$ and $V^N_i=f(A_i^N)$ be the features of the realized and non-realized treatments. We also write $V_i(w)\equiv f(A_i(w))$ for $w\in\{0,1\}$, the features of the two potential treatments. This function may be one-to-one, but in many cases of interest there are multiple elements of ${\cal A}$ mapping onto the same features.  These features capture the fact that the treatments space may be very large, and that in practice we may wish to reduce it to a low-dimensional set of features of the treatment.
 
We observe for all units the quadruple $(Y_i,V_i,V_i^N,X_i)$. We refer to this as a {\it causal observation}, based on a {\it micro experiment} (an experiment of size one). 
It is  an experiment of size one because no other unit is assumed to have the same pair of potential treatments, or even the same pair of feature values $(V_i,V^N_i)$.

\begin{example}[{\sc LLM-Based Pitches}] We use the following as our running example in this discussion. The units, $i=1,\ldots,N,$ are respondents who are shown a pitch $A_i$  to solicit a loan approval on a peer-to-peer lending platform. The outcome $Y_i$ is the approval or not of the loan.
Targeting the specific  respondent an LLM creates set of pitches ${\cal A}_i$ consisting of two or more pitches. These pitches are all unique, and can vary in terms of tone, content, and information about the applicant. 
We extract features $V_i$ of these pitches, including concreteness of the pitch, whether it has guilt appeal, whether it emphasizes impact, whether it emphasizes a story about the loan applicant, and the style of the appeal. The targeting of the respondent by the LLM involves taking into account preferences of the respondent for concreteness of the pitch, or their susceptibility to guilt appeal. This mimics the way in which recommender systems target recommendations based on individual preferences.
One of the elements of the set of pitches ${\cal A}_i$ is selected at random
to be sent to a respondent, with this pitch denoted by $A_i$, and the pitch not selected denoted by $A_i^N$, with corresponding features $V_i$ and $V_i^N$. The respondent then recommends approval  or not of the loan application.
The features of the pitch that is shown are denoted by $V_i$, and the features of the pitch that is not shown are denoted by $V_i^N.$ The approval decision is the outcome, denoted by $Y_i\in\{0,1\}.$ The data are the quadruple $(Y_i,V_i,V_i^N,X_i),$ where $X_i$ may include characteristics of the respondent. 
\end{example}

\begin{example}[{\sc Search Queries}]
Suppose individuals are searching for short term rentals on an online marketplace for rental properties. The treatment set ${\cal A}$ is the set of all properties that may be shown, which may be very large. Individuals put in a search query. Based on their query and  the individual's background, including their rental history, an algorithm selects a set consisting of two properties to potentially show this individual, denoted by ${\cal A}_i=\{A_i(0),A_i(1)\}\subset{\cal A}$. The algorithm then randomly selects one of these two properties to show to the individual, $A_i\in{\cal A}_i$, with the property not shown denoted by $A^N_i\in{\cal A}_i/\{A_i\}$. Finally the individual decides to rent or not the property shown, denoted by $Y_i\in\{0,1\}$. For individual $i$ we observe the  rental choice $Y_i$,  characteristics $X_i$ (these may be a subset or superset of the individual's background that went into the algorithm that selected the unit-specific treatment set ${\cal A}_i$). We also observe some features of both the property shown and the property selected but not shown, $V_i$ and $V_i^N.$ These features may consist of measures such as the per-night rental price of the property, the location, and the average reviews, but these features do not necessarily capture all relevant features of the properties that matter to the individual making the rental choice. In addition, the features of the two selected properties may reflect the individual's preferences, {\it e.g.,} sensitivity to price and quality.
\end{example}

There are five critical features of our  set up that differs from the conventional set up in the causal inference literature.

\noindent{\it 1. There are not necessarily multiple  units with the same treatment.} In the standard experimental set up we have populations with multiple units  randomly assigned to one out of a  set of treatments of low cardinality, with the set of treatments common to all these units, ${\cal A}_i={\cal A}$ for all $i$. For example, all units may either receive aspirin or not, so that ${\cal A}_i={\cal A}=\{{\rm aspirin},{\rm no-aspirin}\}$ for all $i$. As a result we have in the traditional setting many units exposed to the same treatment. In contrast, in our setting the set of potential treatments ${\cal A}_i$ may be unique to unit $i$, and the cardinality of the full set of potential treatments, ${\cal A}\equiv \cup_i {\cal A}_i,$ may be very large, and in fact may be large relative to the sample size. Consequently we may have few, or even no, pairs of units $i$ and $j$ with the same treatment set ${\cal A}_i={\cal A}_j,$ motivating the label micro experiment.

\noindent{\it 2. Distinction between treatment and assignment.} There is a distinction in our case between the assignment $W_i$ and the treatment $A_i.$ This is related to Rubin's SUTVA (stable unit treatment value assumption) which includes the condition that there are no versions of the treatment \cite{rubin1980randomization}. Here we explicitly allow for versions of the treatment. Even in the aspirin example this may be relevant if we are concerned about heterogeneity in the aspirin tablets, {\it e.g.,} the amount of the active ingredient.

\noindent{\it 3. Distinction between treatments and features of the treatments.} There is also a difference between the observed features of the treatment, $V_i$, and the treatment itself, $A_i$.
Although we may observe the actual treatment $A_i$ and the non-realized treatment $A_i^N$, we do not use this information directly. One reason is that the pair of treatments may be unique to a particular unit. In the running example each pitch (treatment) may be unique. In the rental property example there may be too many properties to be able to estimate the effect of a particular property on rental choices. The potential outcomes $Y_i(a)$ may be associated with unobserved features of the treatment. In many analyses it is assumed that the observed features of the treatment fully determine the treatment itself, or that the unobserved features of the treatment are independent of the potential outcomes, but we do not make such assumptions here. The causal interpretation in our analyses does not rely on observing the full set of relevant features of the treatment. Whether we observe the full set of relevant features does affect the interpretation of our estimand, and thus its relative importance, but not its causal nature.
The features can also be extracted from the treatments using algorithms.

\noindent{\it 4. Observability of features of the treatment not received.} The most critical aspect of our set up is that we observe  for each unit $i$ not only the features of the 
realized treatment, $V_i$, but also the features of the non-realized treatment, $V^N_i$, the 
features of the alternative treatment that the unit could have been, but ultimately was not exposed to. 
Such observability is  implicit in many studies in causal inference where each unit has the same potential treatment set ${\cal A}_i$ but it is of the essence here in the context of the heterogeneity in the unit-specific treatment sets.
In conventional experiments with two common treatment levels, say aspirin or no-aspirin with all identical tablets, we implicitly know the treatment, and thus the features of the treatment, each unit was not exposed to. Obviously the features of a treatment the unit was not exposed to cannot have a causal effect on the outcome. However, it can reflect on the assignment mechanism, and in combination with the randomization within the set ${\cal A}_i$ it can eliminate the selection problem.

\noindent{\it 5. Non-observability of the treatment assignment $W_i$.} Finally, we do not observe the treatment indicator $W_i$, only features of the treatment  received and the treatment not received, $V_i$ and $V_i^N$. This is unusual in the causal inference literature, and bears some discussion. One can think of the standard case as the special case where a single observed binary feature of the treatment captures all that matters about the treatment. Note that the assignment $W_i$ cannot be inferred from the observed quadruple $(Y_i,V_i,V_i^N,X_i)$.

\begin{remark} It is instructive to see how the set up relates to
a conventional randomized experiment. Suppose the treatment  corresponds to taking or not  taking  a drug. We can capture that in the current framework by assuming $A_i(w)=w$, for all $w\in\{0,1\}$ and all units. This implies that $A_i=W_i$ and $A_i^N=1-W_i$. Moreover, the observed feature of the treatment is the treatment itself, $V_i=A_i=W_i$ and $V_i^N=A_i^N=1-W_i$. We then  observe for all units the quadruple $(Y_i,W_i,1-W_i,X_i)$, or, equivalently, the triple $(Y_i,W_i,X_i)$. In this special case we trivially know the treatment not received because the unit-specific treatment sets are identical for all units. The two key differences between a conventional randomized controlled trial and the setting we study is that $(i)$ the treatment sets differ by unit, and $(ii)$ the cardinality of the set of potential treatments is potentially high.
\end{remark}

\begin{remark}
Many conventional causal analyses view the features of the treatment $V_i$ as sufficient for the treatment itself. In other words, they do not distinguish between the treatment itself and features thereof. They postulate the existence of potential outcomes indexed by values of these features, $Y_i(v)$.
This implicitly assumes that $Y_i(a)=Y_i(a')$ for all treatments $a$ and $a'$ such that the associated features are identical, $f(a)=f(a')$.
They also assume that conditional on the characteristics $X_i$ the assignment to treatment is unconfounded, $V_i\indep Y_i(v)\ |\ X_i$, for all $v$. Such unconfoundedness property is not implied by the randomization within the set of potential treatments, and we do not make such an assumption. This is the reason that in our setting the conditional expectation $\mathbb{E}[Y_i|V_i=v,X_i=x]$ does not have a causal interpretation as a function of $v$.

\end{remark}

\begin{remark}
The non-realized features $V^N_i$ play a crucial role in our analysis. This role is very different from that of the features of treatments that are not in the unit-specific treatment set ${\cal A}_i$. The non-realized features are informative about this unit-specific treatment set, and thus indirectly about the potential outcomes $Y_i(a)$. This helps address the {\it selection problem} arising from the fact that the realized features $V_i$
may be associated with the potential outcomes 
$Y_i(a)$.

The negative sampling literature, {\it e.g.,}
\cite{he2016vbpr} studies settings where it is not known with certainty whether a non-realized treatment could have been realized or not. That is, for each treatment $a\in{\cal A}/\{A_i\}$ it is not known whether it is an element of the set of treatments that could have been realized, ${\cal A}_i/\{A_i\}$, or an element of the set of treatments that could not have been realized, ${\cal A}/{\cal A}_i.$ Whereas we assume that this distinction is observed, the negative sampling literature assumes that unit-level covariates can help approximate the set ${\cal A}_i$.
Key is that the set ${\cal A}_i/\{A_i\}$ is informative about selection, whereas the set 
${\cal A}/{\cal A}_i$ is not.
\end{remark}

\begin{remark}
Suppose we have data from a number of experiments, each with different treatments. To be specific, consider a set of search engine experiments where various algorithms are investigated. These algorithms may vary by how much weight they put on new potential search results. In each experiment a randomly selected set of search queries  is assigned to a new search algorithm, whereas the remainder of the search queries is assigned to the old algorithm. We could for each search query records characteristics of the query $X_i$, features of the algorithm assigned to the query ({\it e.g.,} the weight it puts on particular aspects of the potential answers shown, features of the algorithm not assigned to the query, and the outcome, say whether the individual clicks on the top result. The particular problem of combining different experiments has been studied previously in \cite{hartford2017deep}, who suggest using the identity of the experiment as an instrument (in the econometrics terminology) for the feature of the treatment of interest.
\end{remark}

Our discussion so far  focuses on the case with two potential treatments. This is solely for expositional reasons. In general there may be a set of potential treatments ${\cal A}_i$ with cardinality larger than 2, and the cardinality may vary by unit. We discuss this in Section \ref{section:multiple_treatments}. We also focus on the case with random assignment to elements of the set of potential treatments with equal probability. This is again for expositional reasons. It is conceptually straightforward to extend our results to the case with known probabilities that differ from $1/2,$ or even to unknown but consistently estimable probabilities,
 and we discuss such generalizations in Section \ref{section:nonequal}.

Next we define the unit-level causal effects. 
\begin{definition}{\sc (The Causal Effect of the Treatment Received)}
\begin{equation}\tau_i\equiv Y_i-Y_i^N
\end{equation}
is the stochastic unit-level causal effect for  unit $i$ of the treatment received relative to the alternative treatment.
\end{definition}
$\tau_i$  is the effect, for unit $i$, of being exposed to treatment they received $A_i$ relative to the treatment they were not exposed to, $A_i^N.$
This definition of $\tau_i$ is different from the conventional one of defining the treatment effect as the difference between the treated and control potential outcomes, in this case $\tilde\tau_i=Y_i(A_i(1))-Y_i(A_i(0)).$
Because we do not observe the assignment $W_i$, and in fact the labeling of the treatments as $A_i(0)$ and $A_i(1)$ is arbitrary, the standard definition is not meaningful here.

These causal effects are {\it doubly heterogeneous}. They are heterogeneous by unit. This is standard in causal inference. They are also, and this is not standard, heterogeneous by the pair of treatments that may be offered to the unit. In the pitch example the pair of potential pitches shown may be relatively concrete for some individuals,   relatively emotional for other individuals, or have more guilt appeal for yet other customers. As a result the micro experiments for each unit can correspond to distinct treatments.

One of the main questions we address in paper is what meaningful averages of these unit level effects we can learn  from a large sample of the causal quadruple $(Y_i,V_i,V_i^N,X_i)$ in this setting.
The first thing to note is that
unconditional averages of the unit-level causal effects $\tau_i$ are not meaningful because the treatments are not ordered in a systematic way, and so the unconditional average is zero by construction.

\begin{lemma} Suppose random assignment (Assumption \ref{ass:randomization}) holds. Then
\[ \mathbb{E}[\tau_i|A_i(0),A_i(1),Y_i(A_i(0)),Y_i(A_i(1)),X_i]=0.\]    
\end{lemma}

Instead we focus on conditional or weighted averages of the $\tau_i$. These weights depend on
functions of the potential treatment features and possibly also on unit-level characteristics. We refer to these averages as {\it conditional treatment-averaged causal effects}, or C-TACEs, which we define formally in the next section.

\section{Estimands, the Treatment Characterizing  Function, and an Identification Result}

Let $h:\mathbb{R}^d\times \mathbb{R}^d\mapsto \mathbb{R}^p$ be an anti-symmetric function, so that $h(v,v')=-h(v',v)$. A natural example is the difference $h(v,v')=v-v'.$ We call this the {\it treatment characterizing function} or TCF.
The function $h(v,v')$ plays a crucial role in our analysis.  It defines the direction in which we characterize the treatment, and implicitly defines the treatment. It is a substantive choice by the researcher, and a critical one.

The TCF defines subpopulations
${\cal P}_h(\delta)\equiv \{i:h(V_i,V_i^N)=\delta\}$. Suppose $V_i$ takes on two values, 0 and 1, and $h(v,v')=\mathbf{1}[v=1,v'=0]-\mathbf{1}[v=0,v'=1]$. Then ${\cal P}_h(1)$ is the subpopulation of units with $V_i=1,V^N_i=0$, and
${\cal P}_h(-1)$ is the subpopulation of units with $V_i=0,V^N_i=1$. The key insight is that the two subpopulations ${\cal P}_h(1)$ and ${\cal P}_h(-1)$, and more generally the subpopulations  ${\cal P}_h(\delta)$ and ${\cal P}_h(-\delta)$,
are comparable by the randomization.

Given the TCF we can define our basic estimand, the {\it Conditional Treatment-Averaged Causal Effect} (C-TACE).
\begin{definition}
{\sc (Conditional Treatment-Averaged Causal Effects)}
\\
The C-TACE for a given TCF $h(v,v')$ and a given feature difference $\delta$ is
\[ \tau_h(\delta)
\equiv \mathbb{E}[\tau_i|h(V_i,V_i^N)=\delta].\]
\end{definition}
In the previous example with a single binary feature $V_i$, $\tau_h(1)$ is the average treatment effect for units who were given the treatment with feature  $V_i=1$ and whose alternative, non-realized treatment had feature $V^N_i=0$. 

Our first and most basic identification result is the following.
\begin{theorem}
Suppose Assumption \ref{ass:randomization} holds. Then the C-TACE $\tau_h(\delta)$ is identified as it can be expressed in terms of the joint distribution of $(Y_i,V_i,V_i^N,X_i)$ as
\[
\tau_h(\delta)=\mathbb{E}[Y_i|h(V_i,V_i^N)=\delta]-\mathbb{E}[Y_i|h(V_i,V_i^N)=-\delta].\]

\end{theorem}

\section{The Pseudo Experiment}\label{sec:pseudo}

The causal quadruple $(Y_i,V_i,V_i^N,X_i)$ is unusual because it does not represent data from a randomized experiment in a standard form.
By the standard form of a randomized experiment we mean an experiment where we observe the triple $(Y_i,W_i,X_i)$ consisting of the treatment $W_i$, the realized outcome $Y_i$, and possibly some  pretreatment variables $X_i$, and where the potential outcomes are independent of the treatment by virtue of the original randomization.

Here we show how we can transform the observed quadruple $(Y_i,V_i,V_i^N,X_i)$ to  data that resemble those from a standard randomized experiment. We refer to this as the {\it pseudo experiment}.
This idea of analyzing the data {\it as if} they were generated by this pseudo  experiment that was not actually performed, but that was implied by the experiment that was performed, echoes the discussion in \cite{athey2018exact}.
There are in fact many ways in the current setting to transform the data to acquire such properties. A very simple way to do this would be to assign the value of a coin flip to each unit and label that the treatment. Such a construction would be meaningless because the average treatment effect for the overall population, as well as the average for  any subpopulation, would by construction be zero. Our specific construction ties the treatment assignment to features of the treatment in such a way that it defines potentially non-zero, and as we argue, meaningful average treatment effects.

Given  the treatment characterizing function (TCF) $h(v,v')$  we  define the three-valued {\it pseudo treatment} $D_i^h$ as
\begin{equation}\label{def_D} D_i^h\equiv 
\left\{
\begin{array}{rl}
1\hskip1cm & {\rm if}\ h(V_i,V_i^N)>0,\\
0\hskip1cm &  {\rm if}\ h(V_i,V_i^N)=0,\\
-1\hskip1cm &  {\rm if}\ h(V_i,V_i^N)<0.
\end{array}\right.
\end{equation}
The indexing of the indicator $D_i^h$ by the TCF $h(\cdot,\cdot)$ stresses the dependence of the definition of the treatment on the TCF.
Note also that $D^h_i$ takes on three values, $\{-1,0,1\}$. 
Units with $D_i^h=0$ play a different role from the other units. Such units would not experience variation in the sign of $h(v,v')$ from changing the the result from the randomization. They are similar to units with propensity score values equal to zero or one in conventional causal analyses.
Similar to the way such units are dropped in conventional causal analyses, we condition in the analyses on $D_i^h\neq 0,$ to account for this.

We also define the {\it pseudo potential outcomes}
\[ V^h_i(-1)=
\left\{
\begin{array}{ll}
V_i(0)\hskip0.1cm & {\rm if}\ h(V_i(0),V_i(1))<0,\\
V_i(1) & {\rm if}\ h(V_i(0),V_i(1))\geq 0,
    \end{array}
\right.
\quad
  V^h_i(1)=
\left\{
\begin{array}{ll}
V_i(1)\hskip0.1cm & {\rm if}\ h(V_i(0),V_i(1))<0,\\
V_i(0) & {\rm if}\ h(V_i(0),V_i(1))\geq 0,
    \end{array}
\right.
\]
\[  V^h_i(0)=(V_i(0)+V_i(1))/2,\]
and
\[ Y^h_i(-1)=
\left\{
\begin{array}{ll}
Y_i(A_i(0))\hskip0.1cm & {\rm if}\ h(V_i(0),V_i(1))<0,\\
Y_i(A_i(1)) & {\rm if}\ h(V_i(0),V_i(1))\geq 0,
    \end{array}
\right.
\quad
  Y^h_i(1)=
\left\{
\begin{array}{ll}
Y_i(A_i(1))\hskip0.1cm & {\rm if}\ h(V_i(0),V_i(1))<0,\\
Y_i(A_i(0)) & {\rm if}\ h(V_i(0),V_i(1))\geq 0,
    \end{array}
\right.
\]
\[  Y^h_i(0)=(Y_i(0)+Y_i(1))/2.\]
\begin{remark}
Note that although we define all six of these potential outcomes, including potential outcomes for the case where $h(V_i(0),V_i(1))=0$ (and thus $D^h_i=0$), we condition in the definition of estimands on $D_i^h\neq 0$ so that the actual definitions in this case are immaterial.
In practice with discrete $V_i$ it is likely there will be a substantial fraction of units with $D_i^h=0.$
\end{remark}

The pseudo treatment $D^h_i$ and the corresponding pseudo potential outcome inherit key properties from the original experiment, at least within the subpopulation with $D^h_i\neq 0$, as captured in the following Lemma.

\begin{lemma}\label{lemma_exp}{\sc (Pseudo Experiment)}
$(i)$ In the subpopulation with $D_i^h\neq 0$ the realized outcomes correspond to the potential outcomes associated with the treatment received,
\[ V_i= V^h_i(D_i^h)=
\left\{
\begin{array}{ll}
 V^h_i(-1)\hskip0.1cm & {\rm if}\ D^h_i=-1,\\
V^h_i(1) & {\rm if}\ D^h_i=1,
    \end{array}
\right.
\quad
 Y_i= Y^h_i(D_i^h)=
\left\{
\begin{array}{ll}
Y^h_i(-1)\hskip0.1cm & {\rm if}\ D^h_i=-1,\\
 Y^h_i(1) & {\rm if}\ D^h_i=1.
    \end{array}
\right.
\]
$(ii)$
Suppose  random assignment (Assumption     \ref{ass:randomization}) holds. Then  $D_i^h$ is independent of the potential outcomes conditional on $D_i^h\neq 0$: 
\[ D_i^h \ \indep \left.\Bigl( V^h_i(-1), V_i^h(1), Y^h_i(-1), Y^h_i(1),X_i\Bigr)\ \right|\  D^h_i\neq 0.\]
\end{lemma}
\begin{remark}
Lemma \ref{lemma_exp}   implies that we can can view the triple $(Y^h_i,D^h_i,X_i)$, within the subpopulation with $D_i^h\neq 0$ as data from a regular experiment with treatment $D^h_i$, outcome $Y_i^h$ and pretreatment variables $X_i$. 
\end{remark}
\begin{remark}
Part $(ii)$ of Lemma \ref{lemma_exp} implies that we can we can treat the pair $( V^h_i(-1), V^h_i(1))$ as pretreatment variables.
Note that this is not true for the pair $(V_i,V_i^N)$. The realized feature $V_i$ is a post-treatment outcome, and conditioning on it is {\it not} generally justified by the randomization of $W_i$ (and the same is true for the non-realized feature $V^N_i)$.
\end{remark}

\begin{lemma}\label{lemma_g}
Suppose that $g(v,v')$ is a symmetric function so that $g(v,v')=g(v',v)$. Then 
\[ D_i^h \ \indep \left.\Bigl( Y^h_i(-1), Y^h_i(1),X_i,g(V_i,V^N_i)\Bigr)\ \right|\  D^h_i\neq 0.\]
\end{lemma}

Let us first consider a simple case. 
In the LLM pitch example, suppose there are two types of pitches, concrete pitches and less concrete pitches. $V_i$ is an indicator  the pitch is concrete $(V_i=1)$ or not $(V_i=0)$. Let $h(v,v')\equiv \mathbf{1}\{v,v'\in\{0,1\}\}\times(\mathbf{1}\{v=1\}-\mathbf{1}\{v'=1\})$, equal to 1 if the first pitch is concrete and the second one is not concrete, equal to -1 if the first pitch is not concrete and the second one is, and zero in all other cases (which includes  the cases where the pitches are of the same type).
Then
\[ D^h_i=h(V_i,V_i^N).\]
Conditioning on $D_i^h\neq 0$ means conditioning on the subpopulation of units where $V_i\neq V_i^N$, that is, the subpopulation with units for whom the selected set of pitches vary in concreteness. The pseudo treatment is an indicator for the pitch shown being a concrete pitch and the pitch not shown being a not-concrete pitch.

Now define
\[ \tau_h\equiv\mathbb{E}[Y_i|h(V_i,V_i^N)>0]-\mathbb{E}[Y_i|h(V_i,V_i^N)<0]
.\]
\begin{lemma}\label{lemma_simpleh}
If $h(v,v')\equiv 2\mathbb{1}_{v-v'>0}-1$, then
\[ \tau_h=\mathbb{E}[\tau_i|V_i>V_i^N]=-\mathbb{E}[\tau_i|V_i<V_i^N].\]
\end{lemma}
$\tau_h$ is a conditional average of the unit-level causal effects. The particular conditioning is subtle. The averaging is over all units where the offered pitch was more concrete than
the non-offered pitch.

Note that this is {\it not} the causal effect of making the pitch more concrete. The features  of all of the pitches are fixed in this analysis. Instead the causal effect is based on looking at the population of all units in  all the micro experiments. In a subset  of these micro experiments there was random assignment between two pitches, with one of the pitches more concrete than the other. These pitches likely also have different values for  other features, associated with concreteness, such as more focus on stories, or more guilt appeal. If $\tau_h$  is positive, the interpretation is that if, instead of randomly choosing between the two currently selected pitches for each individual, we always offer the more concrete one of the two (with its associated other features), the likelihood of the individual recommending funding the loan application increases. This is a valid causal statement, given the assumption of random selection of the offered pitch within the set of selected pitches, but a subtle one.
If the {\it only} feature that individuals care about were concreteness, then the causal effect is also directly a measure of the causal effect of increasing the concreteness. However, this is unlikely with just concreteness as the feature. It could be more plausible if additional features such as story focus, or impact, or guilt appeal were also included.

We can generalize the result in Lemma \ref{lemma_simpleh}. Let $g:\mathbb{R}^d\times \mathbb{R}^d\mapsto \mathbb{R}^q$ be a symmetric function, so that $g(v,v')=g(v',v)$.  For example, $g(v,v')=(v+v')/2$, or $g(v,v')=|v-v'|$. Then
define
\[\tau_{h,g}(x,\gamma)\equiv \mathbb{E}\left[\tau_i\left|X_i=x,h(V_i,V_i^N)>0,g(V_i,V_i^N)=\gamma\right.\right]
\]

\begin{theorem}
Suppose Assumption \ref{ass:randomization} holds. Then 
\[\tau_{h,g}(x,\gamma)=
\mathbb{E}\left[2 D^h_iY_i
\left|X_i=x,g(V_i,V_i^N)=\gamma,D^h_i\neq 0\right.\right]\]
\[=\mathbb{E}\left[Y_i\left|X_i=x,D^h_i=1,g(V_i,V_i^N)=\gamma\right.\right]
-\mathbb{E}\left[Y_i\left|X_i=x,D^h_i=-1,g(V_i,V_i^N)=\gamma\right.\right]
\]

\end{theorem}

\section{Nonequal Assignment Probabilities}\label{section:nonequal}

In this section we extend the results to the case with non-equal assignment probabilities.
The starting point is now the octuple $(W_i,A_i(0),A_i(1),Y_i(0),Y_i(1),P_i(0),P_i(1),X_i)$, where $P_i(w)=\pr(W_i=w)$, so that $P_i(0)+P_i(1)=1$. We observe the quintuple $(V_i,V^N_i,Y_i,P_i,X_i)$ where $P_i=P_i(W_i)$ the probability of the treatment received. We can now define the pseudo experiment given an anti-symmetric function $h(\cdot,\cdot)$ as
\begin{equation} D_i^h\equiv 
\left\{
\begin{array}{rl}
1/(2P_i)\hskip1cm & {\rm if}\ h(V_i,V_i^N)>0,\\
0\hskip1cm &  {\rm if}\ h(V_i,V_i^N)=0,\\
-1/(2(1-P_i))\hskip1cm &  {\rm if}\ h(V_i,V_i^N)<0.
\end{array}\right.
\end{equation}
Now the data for the pseudo experiment are $(Y^h_i,D^h_i,V^h_i,V^{N,h}_i,P_i,X_i)$.

\begin{theorem}
Suppose Assumption \ref{ass:randomization} holds. Then 
\[\tau_{h,g}(x,\gamma)=
\mathbb{E}\left[2 D_i^hY_i
\left|X_i=x,g(V_i,V_i^N)=\gamma,D^h_i\neq 0\right.\right].
\]
\end{theorem}

\section{Multiple Treatment-Characterizing-Functions}\label{sec:multi-tcf}

So far we have focused on the scalar TCF case, where the function $h(v,v')$ maps the product of the feature space into the real line. A scalar TCF leads to a scalar pseudo treatment $D^h_i$ through the definition in (\ref{def_D}). However, the TCF is a researcher choice, and one may wish to consider multiple TCFs. In the running example and empirical illustration, one choice for $h$ is the difference in concreteness of the pitch, let's call this $h_1(v,v')$. A second choice could be the difference in level of guilt in the pitch, denoted by $h_2(v,v')$. Define the corresponding pseudo treatments,
\[ D_i^{h,1}\equiv 
\left\{
\begin{array}{rl}
1\hskip1cm & {\rm if}\ h_1(V_i,V_i^N)>0,\\
0\hskip1cm &  {\rm if}\ h_1(V_i,V_i^N)=0,\\
-1\hskip1cm &  {\rm if}\ h_1(V_i,V_i^N)<0,
\end{array}\right.
\qquad
 D_i^{h,2}\equiv 
\left\{
\begin{array}{rl}
1\hskip1cm & {\rm if}\ h_2(V_i,V_i^N)>0,\\
0\hskip1cm &  {\rm if}\ h_2(V_i,V_i^N)=0,\\
-1\hskip1cm &  {\rm if}\ h_2(V_i,V_i^N)<0.
\end{array}\right.
\]
For ease of exposition let us continue to assume that the assignment to the two pitches is completely random with probability 1/2. As before this implies that
$\pr(D^{h,m}_i=1|D^{h,m}\neq 0)=\pr(D^{h,m}_i=-1|D^{h,m}\neq 0)=1/2$, for both $m=1,2$. However, the two pseudo treatments are potentially correlated.

In this setting, what can we learn about marginal and conditional effects of the two pseudo treatments?
Part of the challenge is that we have a single randomization, which limits how much we can learn from the data.

Lemma \ref{lemma_simpleh} implies that we can
learn the two marginal causal effects
\[ \tau_{h,1}=\mathbb{E}[\tau_i|h_1(V_i,V_i^N)>0]
=\mathbb{E}[Y_i|D^{h,1}_i=1]-\mathbb{E}[Y_i|D^{h,1}_i=-1]
,
\]
and
\[\tau_{h,2}=\mathbb{E}[\tau_i|h_2(V_i,V_i^N)>0]
=\mathbb{E}[Y_i|D^{h,2}_i=1]-\mathbb{E}[Y_i|D^{h,2}_i=-1]
.\]
However, it turns out we can only learn average causal effects such as
\[ \mathbb{E}[\tau_i|h_1(V_i,V_i^N)>0,D^{h,2}_i=d]
\]
for $d=0$, but not for $d=-1,1.$ To see this it is useful to put some more structure on our running example. Suppose there are three types of individuals. For type I the two pitches in the unit-specific treatment set ${\cal A}_i$ have features (with the two features being concreteness and guilt appeal) in $\{(-1,-1),(1,1)\}$, type II has features in 
$\{(-1,1),(1,-1)\}$, and type III has features in 
$\{(-1,1),(1,1)\}$. Thus the first type of individuals get pitches that are high in concreteness and high in guilt or low in both, the second type gets pitches that are high in concreteness but low in guilt appeal or low in concreteness and high in guilt appeal, and the third type  gets pitches that vary in concreteness but that are always high in guilt appeal.

First consider the interpretation of $\tau_{h,1}$ and $\tau_{h,2}$. 
For the first type, the average effect of pseudo treatments 1 and 2 are identical, because pseudo treatments $D^{h,1}_i$
and $D^{h,2}_i$ are perfectly positively correlated.  For the second type the two pseudo treatments are perfectly negatively  correlated. For the third type 
$\tau_{h,1}$ measures the effect of concreteness for units with $D_i^{h,2}=0$, and $\tau_{h,2}$ is not defined. The overall effects $\tau_{h,1}$ and $\tau_{h,2}$ are weighted averages of these conditional effects (with $\tau_{h,2}$ being defined conditional on $D^{h,2}_i\neq 0$. If there are only individuals of the first type, $\tau_{h,1}=\tau_{h,2}$, and if there are only individuals of the second type, then
$\tau_{h,1}=-\tau_{h,2}$.

Second, let us consider the claim that we cannot learn
$\mathbb{E}[\tau_i|h_1(V_i,V_i^N)>0,D^{h,2}_i=d]$ for $d=-1,1$. Conditional on $D^{h,2}_i=d$ there is no variation in $D^{h,1}_i$ for individuals of the first type because $D^{h,1}_i$ and $D^{h,2}_i$ are perfectly positively correlated for these individuals. Nor  is there variation in $D^{h,1}_i$ for individuals of the second type because $D^{h,1}_i$ and $D^{h,2}_i$ are perfectly negatively correlated for these individuals. Thus we cannot learn $\mathbb{E}[\tau_i|h_1(V_i,V_i^N)>0,D^{h,2}_i=d]$.

However, we can learn 
$\mathbb{E}[\tau_i|h_1(V_i,V_i^N)>0,D^{h,2}_i=0]$. For units with $D_i^{h,2}=0$ we can estimate  $\mathbb{E}[\tau_i|h_1(V_i,V_i^N)>0,D^{h,2}_i=0]$ through the equality
\[ \mathbb{E}[\tau_i|h_1(V_i,V_i^N)>0,D^{h,2}_i=0]=
\mathbb{E}[Y_i|D^{h,1}_i=1,D^{h,2}_i=0]
-\mathbb{E}[Y_i|D^{h,1}_i=-1,D^{h,2}_i=0],
\]
with both conditioning events having positive probability for the type III individuals.

\section{Multiple Treatments}\label{section:multiple_treatments}

So far we have focused on the case with two treatments, $A_i(0)$ and $A_i(1)$. In many cases there are multiple treatments. Here we discuss the case with one realized treatment and $K$ non-realized treatments, each with potentially different probabilities. Now the data are $(Y_i,V_i,V^{N1}_i,\ldots,V^{NK}_i,P_i,P^{N1}_i,\ldots,P^{NK}_i,X_i)$. The insight is that we can think of this as $K$ micro experiments of size one. We exploit this insight by turning this observation into $K$ causal data points, with the $k$th causal observation of the form
$(Y_i,V_i,V^{Nk}_i,P_i/(P_i+P^{Nk}_i),X_i)$. 

\section{Estimated Assignment Probabilities and Counterfactual Exposures}\label{section:estimated_probs}

The analysis in Section \ref{section:nonequal} takes the assignment probability $P_i$ to be part of the data. In some applications, this may not be observed: production systems rarely log the probability with which a particular item was displayed. What such systems frequently do permit, however, is either the logging of additional items that were considered but not exposed (for example, the set of candidate items selected to be passed to a final ranking algorithm), or the re-execution of the display mechanism at the logged configuration. An LLM generates responses by sampling at a fixed temperature, and the generation step can be run again at the same prompt; in the rental example the ranking algorithm can be re-run on the logged query. In this section we formalize the corresponding data requirements, show that the estimator of Section \ref{section:nonequal} remains valid when the assignment probabilities are estimated by such replays, and characterize the number of replays required. Throughout this section we specialize to a scalar binary feature, $V_i\in\{0,1\}$, with the TCF
\[ h(v,v')=\mathbf{1}\{v=1,v'=0\}-\mathbf{1}\{v=0,v'=1\},\]
as in Section \ref{sec:pseudo}, so that the discordant pairs are the units with $D^h_i\neq 0$. The results extend to any single scalar TCF whose sign can be computed from a replayed exposure.

\subsection{Observability regimes}

We distinguish three data regimes, ordered by how much the analyst observes about the assignment mechanism.

\begin{definition}{\sc (Observability Regimes)}\label{def:regimes}
\begin{itemize}
\item[(I)] {\it Logged-probability data.} For each unit we observe the quintuple $(Y_i,V_i,V^N_i,P_i,X_i)$, with $P_i$ the assignment probability of Section \ref{section:nonequal}.
\item[(II)] {\it Counterfactual-exposure data.} For each unit we observe $(Y_i,V_i,V^N_{i,1},\ldots,V^N_{i,M},X_i)$, where $V^N_{i,1},\ldots,V^N_{i,M}$ are the features of $M$ {\it counterfactual exposures}: items produced by the display mechanism at unit $i$'s configuration that were not exposed to the unit. The case $M=1$ returns the causal quadruple of Section \ref{sec:setup}.
\item[(III)] {\it Replayable-mechanism data.} We observe the causal quadruple $(Y_i,V_i,V^N_i,X_i)$ and, in addition, the analyst can query the display mechanism at unit $i$'s configuration, obtaining further counterfactual exposures on demand. The number of queries per unit is a design choice.
\end{itemize}
\end{definition}

Regime (II) describes datasets in which the counterfactual exposures were logged at the time of the interaction, for example the candidate pitches an LLM generated but did not send, or the runner-up listings a ranker considered. Regime (III) describes the situation where the mechanism itself is available to the analyst after the fact. The two regimes support the same estimators; the difference is that in Regime (III) the analyst chooses the sampling design, including sequentially, which Section \ref{sec:inverse_sampling} exploits.

The identifying condition in both regimes is that the counterfactual exposures are draws from the same stochastic mechanism that produced the realized exposure. Write $q_i$ for the probability that the display mechanism at unit $i$'s configuration produces the high-feature side,
\[ q_i\equiv \pr\Bigl(V_i=1\Bigl|A_i(0),A_i(1),Y_i(A_i(0)),Y_i(A_i(1)),X_i\Bigr),\]
so that on the discordant population the assignment probability of Section \ref{section:nonequal} is $P_i=q_i$ when the high-feature side is realized ($D^h_i>0$) and $P_i=1-q_i$ when the low-feature side is realized ($D^h_i<0$).

\begin{assumption}\label{ass:reproducible}{\sc (Reproducible Randomization)}
Conditional on $\bigl(A_i(0),A_i(1),Y_i(A_i(0)),Y_i(A_i(1)),X_i\bigr)$, the feature indicators of the realized exposure and of the counterfactual exposures,
\[ \mathbf{1}\{V_i=1\},\ \mathbf{1}\{V^N_{i,1}=1\},\ \ldots,\ \mathbf{1}\{V^N_{i,M}=1\},\]
are independent Bernoulli draws with common success probability $q_i$.
\end{assumption}

Assumption \ref{ass:reproducible} plays the role that the observability of $P_i$ played in Section \ref{section:nonequal}: it states that the replays carry the assignment randomization. It is substantive. In Regime (II) it requires that the logged counterfactual exposures were generated by the same process, and not, for instance, by a deterministic runner-up rule; in Regime (III) it requires that the re-executed mechanism matches the production mechanism at the time of exposure (same model version, prompt, and sampling temperature in the LLM example). Section \ref{sec:replay_diagnostic} gives a diagnostic. When the display mechanism exposes its choice probabilities directly, as when an LLM API returns token-level log-probabilities, Regime (I) obtains and no replays are needed; the methods of this section are useful when probabilities are not available but the algorithm can be replayed.

\subsection{Overlap trimming and its relation to ties}\label{sec:trimming}

Under Assumption \ref{ass:reproducible} the natural estimator of $q_i$ in Regime (II) is the sample frequency
\[\hat q_i\equiv\frac{1}{M+1}\left(\mathbf{1}\{V_i=1\}+\sum_{j=1}^M \mathbf{1}\{V^N_{i,j}=1\}\right),\]
which pools the realized exposure with the counterfactual exposures, all $M+1$ indicators being exchangeable. Estimating $q_i$ is not difficult: $\hat q_i$ is unbiased. However, we require estimation of the {\it weights} $1/(2P_i)$ and $1/(2(1-P_i))$ that enter the pseudo treatment in Section \ref{section:nonequal}: the map $p\mapsto 1/p$ is convex, so the plug-in weight is biased upward, with relative bias of order $(1-P_i)/(M P_i)$, and the bias does not average out across units. Define the trimmed population
\[ {\cal P}(\kappa)\equiv\bigl\{i: \kappa\leq q_i\leq 1-\kappa\bigr\},\qquad \kappa\in(0,1/2),\]
and restrict the estimands of Section \ref{section:nonequal} to ${\cal P}(\kappa)$.

Trimming on $q_i$ is the probabilistic counterpart of dropping ties. A tied pair, $D^h_i=0$, is a micro experiment that carries no feature contrast: whichever side is displayed, the unit is exposed to the same feature value, and Section \ref{sec:ties} shows that retaining such pairs dilutes the estimand or worse. A unit with $q_i$ near zero or one is a {\it stochastic tie}: the pair is discordant, but the mechanism essentially always displays the same side, so the micro experiment carries almost no assignment variation. Some users may, for example, be so strongly targeted that they essentially always receive the concrete pitch; for such users the display carries very little usable randomization, and the inverse weight, which grows as $1/q_i$, creates variance. Excluding units based on extreme propensities reflects the same qualitative design decision as excluding the micro experiments that contain no experiment, and both redefine the estimand to the retained population.

Trimming also alleviates the sample-size requirements for $M$. On ${\cal P}(\kappa)$ the relative bias of the plug-in weight is at most $(1-\kappa)/(M\kappa)$, so a tolerance $\varepsilon$ requires only $M\geq(1-\kappa)/(\kappa\varepsilon)$; because extreme probabilities are excluded by design rather than estimated, $M$ need not be large. A first-order bias correction, replacing $1/\hat q_i$ by $(1/\hat q_i)\bigl(1-(1-\hat q_i)/((M+1)\hat q_i)\bigr)$, reduces the bias to order $M^{-2}$ and the requirement to $M\gtrsim\sqrt{(1-\kappa)/(\kappa\varepsilon)}$. The sequential design of the next subsection removes the bias entirely.

\subsection{Inverse sampling for a single feature}\label{sec:inverse_sampling}

In Regime (III) the analyst controls the sampling design, and when a single TCF is the target a sequential design yields an exactly unbiased weight. The approach proceeds as follows: fix an integer $r\geq 1$. For unit $i$ in the discordant, trimmed population, replay the stochastic algorithm repeatedly, classifying the exposed item for each replay by the binary feature, and stopping when the feature value that was actually realized for unit $i$ has occurred $r$ times. Let $T_i$ denote the total number of replays at stopping. Because every replay is classified, no draw is wasted, and $T_i$ follows a negative binomial distribution with parameters $r$ and $P_i$.

\begin{lemma}\label{lemma_inverse_sampling}{\sc (Inverse Sampling)}
Suppose Assumption \ref{ass:reproducible} holds, extended to the replay draws of Regime (III). Then, conditional on $\bigl(A_i(0),A_i(1),Y_i(A_i(0)),Y_i(A_i(1)),X_i\bigr)$ and on $D^h_i\neq0$,
\[ \mathbb{E}\left[\frac{T_i}{r}\right]=\frac{1}{P_i},
\qquad
\mathbb{V}\left(\frac{T_i}{r}\right)=\frac{1-P_i}{r P_i^2}.\]
\end{lemma}

The unbiasedness is exact at every $r\geq1$, in contrast to fixed-$M$ plug-in weights, whose bias vanishes only as $M$ grows; this is the classical property of inverse binomial sampling \citep{haldane1945inverse,girshick1946unbiased}. Substituting the sampled weight into the pseudo treatment of Section \ref{section:nonequal} gives the feasible pseudo treatment
\[ \hat D_i^h\equiv \left\{
\begin{array}{rl}
T_i/(2r) & {\rm if}\ h(V_i,V^N_i)>0,\\
0 & {\rm if}\ h(V_i,V^N_i)=0,\\
-T_i/(2r) & {\rm if}\ h(V_i,V^N_i)<0.
\end{array}\right.\]

\begin{theorem}\label{theorem_feasible}
Suppose Assumption \ref{ass:randomization}, generalized as in Section \ref{section:nonequal}, and Assumption \ref{ass:reproducible} hold, and that the replay draws are independent of the potential outcomes conditional on the configuration. Then, on the population with $D^h_i\neq 0$ and $q_i\in[\kappa,1-\kappa]$,
\[ \tau_{h,g}(x,\gamma)=\mathbb{E}\left[2\hat D_i^h Y_i\,\Bigl|\,X_i=x, g(V_i,V^N_i)=\gamma, D^h_i\neq0\Bigr.\right],\]
and the conditional variance of the feasible moment exceeds that of the infeasible moment $2D^h_iY_i$ by
\[ \mathbb{E}\left[\frac{1-P_i}{r P_i^2}\,Y_i^2\,\Bigl|\,\cdot\Bigr.\right]\ \leq\ \frac{1-\kappa}{r\kappa^2}\,\mathbb{E}\bigl[Y_i^2\bigl|\cdot\bigr.\bigr].\]
\end{theorem}

The theorem separates the two costs of not observing $P_i$. There is no bias cost: the feasible estimator identifies the same estimand as the theorem of Section \ref{section:nonequal}, because the sampled weight is unbiased for the true weight and the replay noise is independent of the outcome. There is a variance cost, which vanishes at rate $1/r$ and is bounded on the trimmed population. In practice, values of $r$ between five and ten keep the inflation to a few percent for bounded outcomes. The expected number of replays for unit $i$ is $r/P_i\leq r/\kappa$, so the per-unit query cost is modest precisely because the trimming of Section \ref{sec:trimming} has excluded the configurations whose probabilities would be expensive to sample.

\begin{remark}
A maximum replay budget per unit, $T_i\leq \bar T$, acts as an implicit trim: units that exhaust the budget before $r$ matches are, with high probability, units with $P_i\lesssim r/\bar T$. Aligning the explicit trim with the budget, $\kappa\approx r/\bar T$, and dropping the censored units keeps the estimand well defined on ${\cal P}(\kappa)$; imputing weights for censored units does not.
\end{remark}

\begin{remark}\label{remark_multifeature}
The design extends to a set of TCFs fixed in advance, using a single replay stream per unit. For each feature $c$ in the set, let $T^{(c)}_i$ be the index of the replay at which the realized value of feature $c$ attains its $r$-th match, and continue collecting counterfactual exposures until every feature in the set has done so. Although the overall stopping rule depends on all the features jointly, each $T^{(c)}_i$ is a function of the feature-$c$ classifications alone and is determined at its own stopping index, regardless of how much longer the stream runs; hence each $T^{(c)}_i$ is exactly negative binomial with parameters $(r,P^{(c)}_i)$, and Lemma \ref{lemma_inverse_sampling} and Theorem \ref{theorem_feasible} apply feature by feature. The weights for different features are computed from a common stream and are therefore correlated within unit, which is immaterial for estimating each C-TACE separately but enters the covariance matrix for joint inference across TCFs. The expected stream length is determined by the feature with the least overlap, $\mathbb{E}[\max_c T^{(c)}_i]\geq r/\min_c P^{(c)}_i$, so the trimming of Section \ref{sec:trimming} should be applied per feature: the trimmed population is feature specific, ${\cal P}_c(\kappa)=\{i:\kappa\leq q^{(c)}_i\leq 1-\kappa\}$, and a unit may be retained for one TCF while trimmed for another.
\end{remark}

\begin{remark}\label{remark_resume}
The same argument accommodates draws collected at different times. A stored set of $M$ counterfactual exposures (Regime (II)) may serve as the beginning of the replay stream. If the $r$-th match of the realized feature value occurs within the stored draws, no further queries are needed; if not, sampling resumes until it does, and $T_i$ remains exactly negative binomial because the i.i.d.\ stream does not depend on when its elements were collected. Likewise, for a feature defined after the original data collection, the stored replays are classified retroactively, provided the raw responses rather than only their feature values were retained, and sampling resumes for the units whose stored draws contain fewer than $r$ matches. Resumption is essential: restricting a new feature's analysis to units whose stored draws happen to contain $r$ matches selects on high $P^{(c)}_i$ and replaces the negative binomial count with a truncated one, reinstating the bias. Resumption also extends Assumption \ref{ass:reproducible} across time, requiring that the stochastic algorithm queried later is the same algorithm that generated the stored draws and the exposures; the diagnostic of Section \ref{sec:replay_diagnostic}, applied to stored versus resumed draws, provides a check. As of 2026, it has been common practice for model providers to maintain access to most, but not all, older models, however some models have been deprecated. Open weight models that are locally hosted can be maintained indefinitely. In this respect, LLMs differ from most production recommendation systems, where model parameters and algorithms may be updated at high frequency, and where resampling later may sacrifice validity.
\end{remark}

\subsection{A diagnostic}\label{sec:replay_diagnostic}

Assumption \ref{ass:reproducible} is testable. Under the assumption, the realized indicator $\mathbf{1}\{V_i=1\}$ is one further draw from the same distribution that generated the replays, so across units the realized exposures must be calibrated to the observed replay frequencies. In any group of units with replay frequency $\hat q_i\approx q$, a fraction $q$ must have realized the high-feature side. A regression of $\mathbf{1}\{V_i=1\}$ on $\hat q_i$ with intercept zero and slope one, or a binned comparison of realized shares against replay frequencies, can be used to test the hypothesis that the replays come from the same distribution.

\section{A Linear Example}\label{sec:linear-example}

The results so far are nonparametric. To make the estimand and the alternatives to it concrete, and to show what each recovers, we specialize the setup of Section~\ref{sec:setup} to a linear data-generating process and compute the probability limit of each estimator in closed form. The derivations are collected in Section~\ref{sec:example-derivations}; the figures are produced by the replication code in the supplement.

\subsection{The data-generating process}\label{sec:dgp-example}

The treatments are pitches drawn from the catalog $\mathcal{A}$; each treatment $a$ has feature vector $f(a)\in\mathbb{R}^d$ whose two coordinates we track are concreteness $f(a)_c$ and story emphasis $f(a)_s$. A respondent $i$ has observed preferences $X_i$, with $X_{i,c}$ the preference for concreteness, and an unobserved preference or trait $\xi_i$. Potential outcomes are generated by a linear response: for any treatment $a$,
\[
\mathbb{E}\!\left[Y_i(a)\mid X_i,\xi_i\right]=m\big(f(a),X_i,\xi_i\big),\qquad
m(v,x,\xi)=\alpha_0+\beta_c v_c+\beta_s v_s+\eta_c x_c+\delta_{cc}\,v_c x_c+\psi\,\xi,
\]
and $Y_i(a)$ is a Bernoulli draw at this mean.

The (offered, non-offered) pair is chosen by a recommender. It assigns each respondent/pitch combination an estimated score (in a recommendation system, the system's own predicted match for the respondent),
\[
U_i(a)=\lambda_{cc}\,f(a)_c\,X_{i,c}+\lambda_{ss}\,f(a)_s\,X_{i,s}+\sigma\,\varepsilon_{ia}.
\]
The recommender then
selects for respondent $i$ the two highest-scoring candidates, $\mathcal{A}_i=\{A_i(0),A_i(1)\}$, with feature vectors $V_i(0)=f(A_i(0))$ and $V_i(1)=f(A_i(1))$. As discussed in Section~\ref{sec:setup}, which of the two selected pitches is displayed is then randomized, independently of the potential outcomes given the offered set, so that $\Pr(W_i=1\mid\mathcal{A}_i,Y_i(A_i(0)),Y_i(A_i(1)),X_i)=P$ with $P=1/2$ in the baseline (Assumption~\ref{ass:randomization}). As in Section~\ref{sec:setup}, $V_i=f(A_i)$, $V_i^N=f(A_i^N)$, $Y_i=Y_i(A_i)$, and we observe the quadruple $(Y_i,V_i,V_i^N,X_i)$.

The example makes two key forces of the general setup concrete. The recommender's scoring rewards through the parameters $\lambda_{cc}$ the match between a pitch's concreteness and the respondent's preference for it, so the offered set $\mathcal{A}_i$ depends on $X_{i,c}$. In other words, which pitches a respondent may see is  correlated with their preferences. Preferences in turn move the outcome whenever $\eta_c+\delta_{cc} V_{ic}\neq0$. What is not confounded is the choice within the unit-specific treatment set. Given $\mathcal{A}_i$, the display $W_i$ is  random. The proposed analysis turns on this split: a selected unit-specific treatment set, and a randomized display within it.

\subsection{The C-TACE Estimator, specialized}\label{sec:withinpair}

The C-TACE estimator is built from the pseudo-experiment of Section~\ref{sec:pseudo}, with the treatment-characterizing function specialized to concreteness, $h(v,v')=v_c-v'_c$, and the three-valued pseudo-treatment $D_i^h$ defined in Equation~\eqref{def_D}. Its identification was established there. Here we evaluate what estimand recovers under the linear response. Write the offered pair's more- and less-concrete pitches as $V_i^{h}(1)$ and $V_i^{h}(-1)$, with concreteness and story gaps  $\Delta_{i,c}=V_{i,c}^{h}(1)-V_{i,c}^{h}(-1)$
and $\Delta_{i,s}=V_{i,s}^{h}(1)-V_{i,s}^{h}(-1)$
, and the transformed outcome as $\tilde Y_i=2D_i^h Y_i$.

Under the linear response the treatment-averaged concreteness contrast for a pair is
\[
m(V_i^{h}(1),X_i,U_i)-m(V_i^{h}(-1),X_i,U_i)=\beta_c\,\Delta_{i,c}+\beta_s\,\Delta_{i,s}+\delta_{cc}\,\Delta_{i,c}\,X_{i,c},
\]
with average
\[\tau=\mathbb{E}[\tilde Y_i|D_i^h\neq 0]=\beta_c\mathbb{E}[\Delta_{i,c}|D_{i}^h\neq 0]
+\beta_s\mathbb{E}[\Delta_{i,s}|D_{i}^h\neq 0]
+\delta_{cc}
\mathbb{E}[\Delta_{i,c}X_{i,c}|D_{i}^h\neq 0]
\]
Because the contrast is taken within a single respondent's selected pair, the respondent-specific terms $\eta_c X_{i,c}$ and $\psi U_i$ enter both terms identically and do not affect the difference; only the within-pair feature gaps $\Delta_i$ remain. Dividing the mean of $\tilde Y_i$ over the discordant pairs by the mean gap $\bar h_c=\mathbb{E}[\Delta_{i,c}\mid\Delta_{i,c}>0]$ expresses the estimand per unit of concreteness. This per-unit of concreteness estimand is $\beta_c$ when no other feature moves with concreteness ($\mathbb{E}[\Delta_{i,s}]=0$) and there is no interaction between the concreteness feature and preference for concreteness ($\delta_{cc}=0$). In general  the per-unit concreteness estimand is equal to $\beta_c$ plus the effect of the features that co-vary with concreteness across the pair, the bundled quantity we write as $\theta=\tau/\bar h_c$.

\subsection{Ties}\label{sec:ties}

A tied pair is one whose two pitches have equal concreteness, $V_{i,c}^{h}(1)=V_{i,c}^{h}(-1)$, so $\Delta_{i,c}=0$, $D_i^h=0$, and $\tilde Y_i=0$. Such a pair is uninformative about the effect of concreteness. Within a tied pair concreteness does not vary, so there is no contrast from which to learn its effect: a pair tied at high concreteness shows the respondent two concrete pitches, a pair tied at low concreteness shows two plain ones, and in neither case does the respondent face a change in concreteness.

The tied pairs are also not one homogeneous group. The marginal effect of concreteness, $\partial m/\partial v_c=\beta_c+\delta_{cc}x_c$ (and, if the response is curved in concreteness, it depends on the level itself), varies across respondents, so the concreteness effect one would want for the pairs tied at high concreteness differs in general from the effect for the pairs tied at low concreteness. Yet within each of these groups concreteness is held fixed, so neither effect is identified from its own group: there is no variation in $v_c$ to identify it. There is no control group inside a tied group.

It is tempting to salvage the tied pairs by comparing across the groups: the mean funding $\mathbb{E}[Y_i\mid\text{tie-high}]$ among the pairs tied at high concreteness against the mean $\mathbb{E}[Y_i\mid\text{tie-low}]$ among those tied at low concreteness. But this is a comparison of systematically different respondents, not a within-pair contrast. Under the recommender's sorting the high-high pairs go to respondents with a high preference for concreteness and the low-low pairs to respondents with a low one, so
\[
\mathbb{E}[Y_i\mid\text{tie-high}]-\mathbb{E}[Y_i\mid\text{tie-low}]=\beta_c\big(\bar v_c^{\,\text{hi}}-\bar v_c^{\,\text{lo}}\big)+\eta_c\big(\bar X_c^{\,\text{hi}}-\bar X_c^{\,\text{lo}}\big)+\dots
\]
carries the preference gap $\eta_c(\bar X_c^{\,\text{hi}}-\bar X_c^{\,\text{lo}})$ alongside any concreteness difference, and is not a causal effect of concreteness.

\begin{figure}[tbp]\centering
\includegraphics[width=0.72\linewidth]{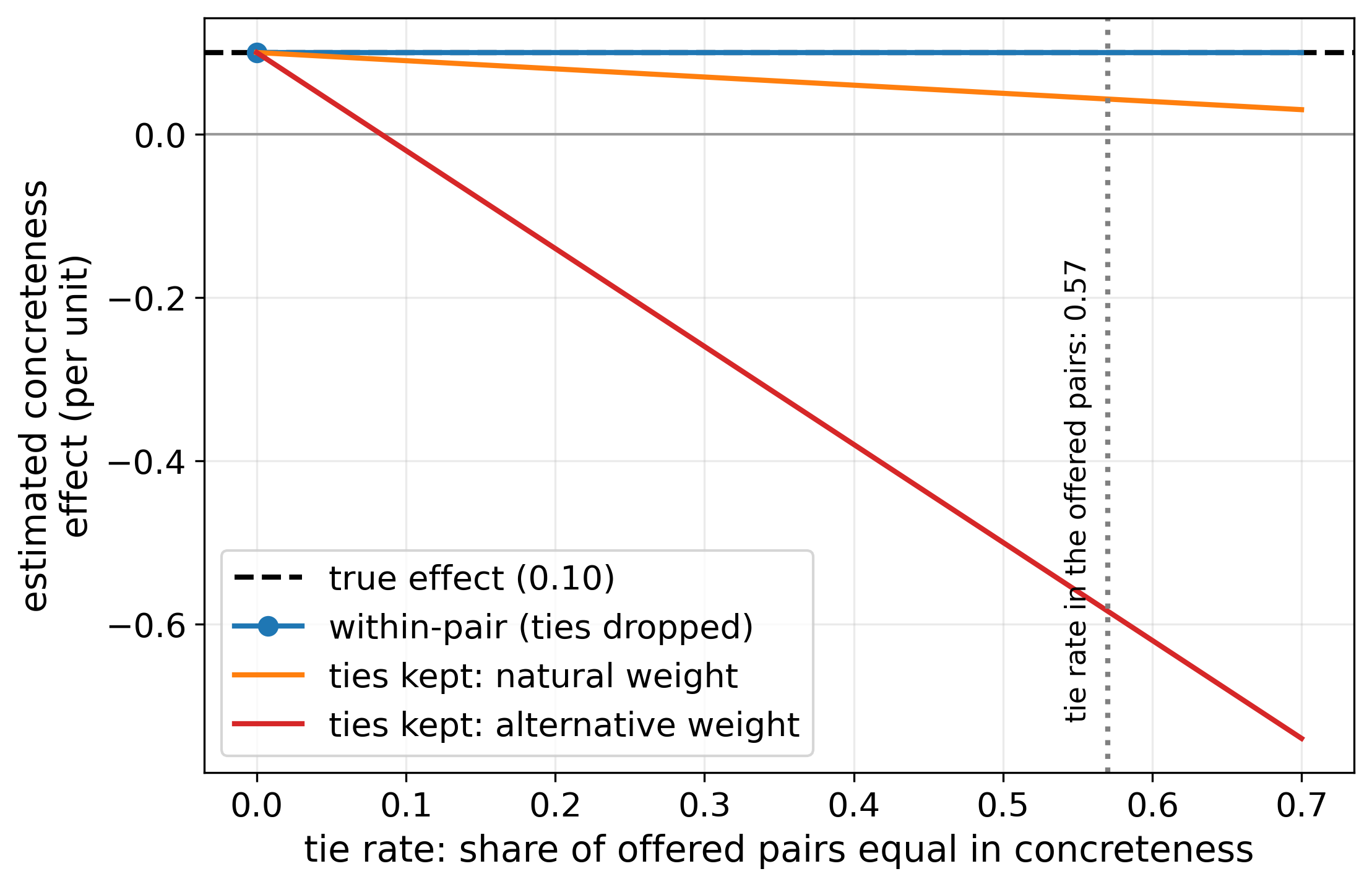}
\caption{Dropping the pairs with equal $v_c$. Estimate against the share of tied pairs $\pi_0$: dropping ties holds the effect at $\beta_c$, keeping them under the natural weight dilutes the estimate toward zero, and under the alternative weight drives it below zero.}
\label{fig:ex-ties}
\end{figure}

Dropping the pairs with equal $v_c$ is therefore part of the identification strategy. If we leave them in the within-pair average they add zeros to the numerator while still counting in the denominator, and with $\pi_0$ the tied fraction the estimator converges to the diluted $(1-\pi_0)\,\tau_h$; a weighting that instead gives each tied pair a nonzero weight, as though its less-concrete member had been displayed, adds a term $-\tfrac{\pi_0}{1-P}\,\mathbb{E}[Y_i\mid\text{tie}]$ that depends only on how often tied pairs are funded and can turn the estimate negative. Figure~\ref{fig:ex-ties} shows the consequence of keeping versus dropping.

\subsection{The Naive Regression}\label{sec:naive}

\begin{figure}[tbp]\centering
\includegraphics[width=0.72\linewidth]{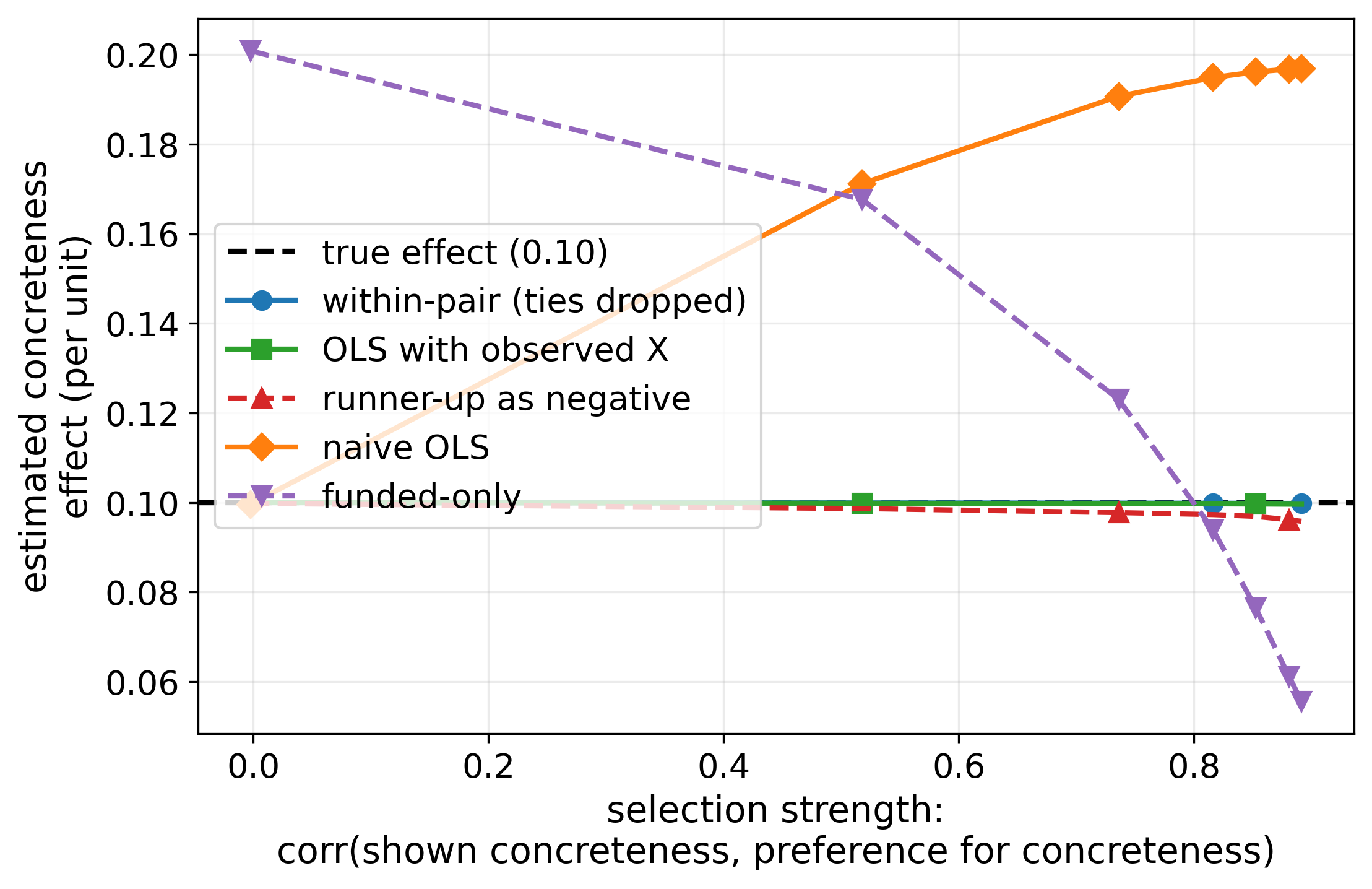}
\caption{Robustness to selection. Probability limits of the estimators as the recommender sorts more strongly on concreteness; the horizontal axis is the correlation between the shown pitch's concreteness and the respondent's preference. The C-TACE estimator, covariate-adjusted regression, and the negative-example regression that uses the randomized runner-up all stay at $\beta_c$; only the naive regression drifts, by $\eta_c b$.}
\label{fig:ex-selection}
\end{figure}

One common analysis uses only the pitch each respondent was actually shown: it regresses the realized funding outcome $Y_i$ on the concreteness of that displayed pitch $V_{i,c}$ across respondents, discarding the non-shown pitch $V_{i,c}^N$ and the pairing. Because each respondent then contributes a single observation and the comparison is across different respondents, it no longer holds the respondent fixed, and the preference is not differenced out. The least squares slope converges to $\operatorname{Cov}(V_{i,c},Y_i)/\operatorname{Var}(V_{i,c})$; using $\mathbb{E}[Y_i\mid V_i,X_i,U_i]=m(V_i,X_i,U_i)$,
\[
\operatorname{plim}\hat\beta^{\text{OLS}}=B_0+\eta_c\,\frac{\operatorname{Cov}(V_{i,c},X_{i,c})}{\operatorname{Var}(V_{i,c})}.
\]
The extra term, $\eta_c\,\operatorname{Cov}(V_{i,c},X_{i,c})/\operatorname{Var}(V_{i,c})$, is the standard omitted-variable bias of a regression that conditions only on the shown pitch: the displayed concreteness is correlated with the respondent's preference, and the preference also affects funding. The naive analysis inherits this familiar confounding. In this example the recommender shows more concrete pitches to respondents who prefer concreteness, so $\operatorname{Cov}(V_{i,c},X_{i,c})>0$; those respondents fund more often, so $\eta_c>0$; and the regression credits concreteness with what is their higher baseline generosity. The bias is the product of how strongly the recommender sorts on concreteness and how much the preference it sorts on matters for funding, and vanishes only if one of the two is zero. Figure~\ref{fig:ex-selection} traces this comparative static as the sorting is turned up.

\subsection{Controlling for the Covariate, and its Probability Limit}\label{sec:controlx}

\begin{figure}[tbp]\centering
\includegraphics[width=0.72\linewidth]{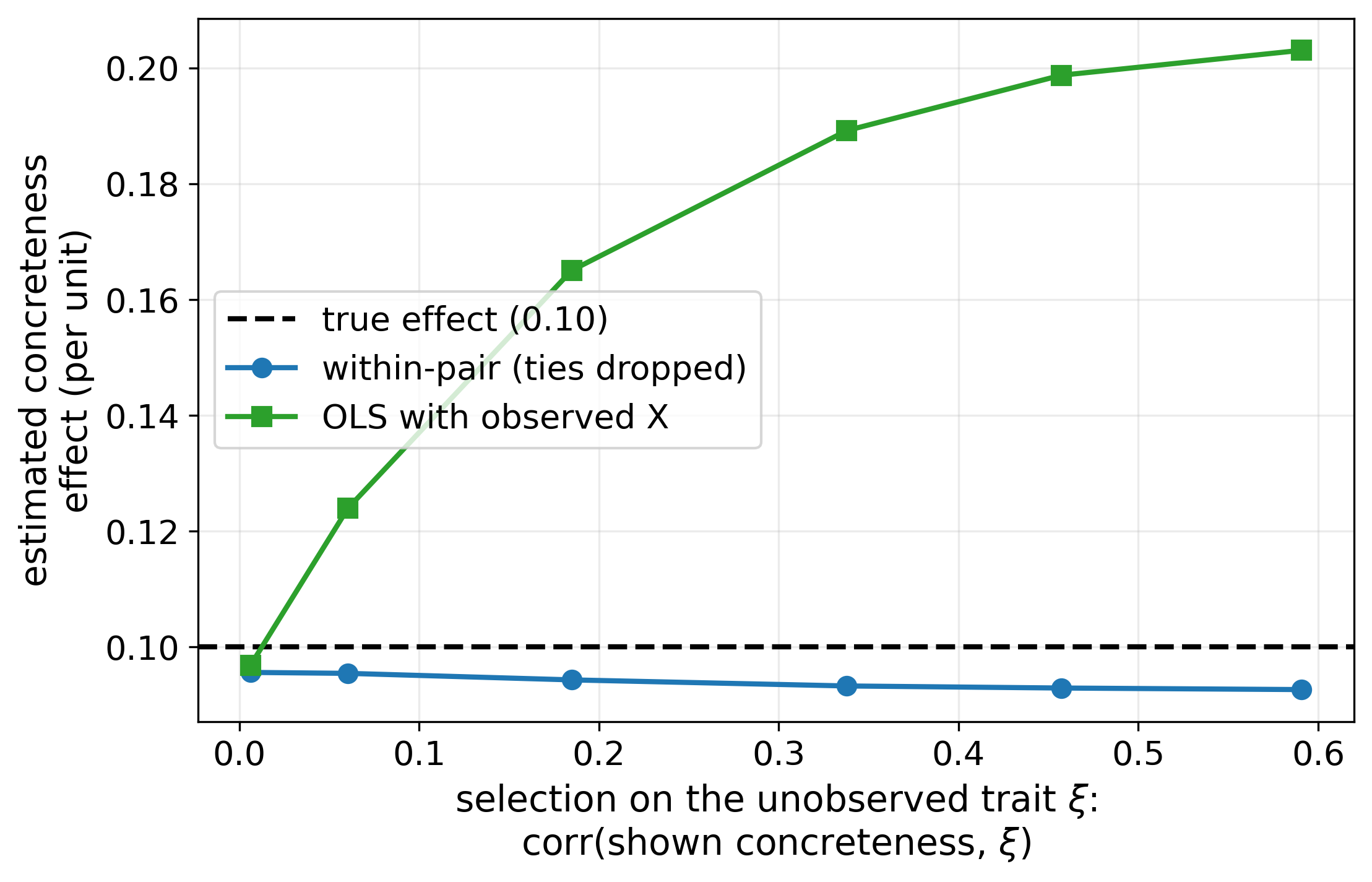}
\caption{An unobserved trait the analyst cannot control for. As the recommender sorts on an unobserved respondent trait that also affects funding, the covariate-adjusted (unconfoundedness) regression drifts away from $\beta_c$, while the within-pair estimate is unchanged.}
\label{fig:ex-unobs}
\end{figure}

The bias in the naive regression comes from the omitted preference, so a natural repair is to include it: regress $Y_i$ on the shown concreteness $V_{i,c}$ and the covariate $X_i$, on the shown sample. This is the unconfoundedness analysis of Section~\ref{sec:unconf}, which treats the shown feature as unconfounded once $X_i$ is held fixed. By the Frisch--Waugh--Lovell theorem the coefficient on $V_{i,c}$ is the slope of $Y_i$ on the part of $V_{i,c}$ orthogonal to $X_i$, so, writing $V_{i,c}^{\perp}$ for that residual,
\[
\operatorname{plim}\hat\beta^{\text{unc}}=\theta+\psi\,\frac{\operatorname{Cov}(V_{i,c}^{\perp},\xi_i)}{\operatorname{Var}(V_{i,c}^{\perp})}.
\]
When the recommender sorts only on observed preferences, $\xi_i$ is irrelevant and the estimator returns the bundled effect $B_0$. The residual term survives only when the offered set depends on a trait $U_i$ that the analyst does not observe and that also moves funding ($\psi\neq0$): controlling $X_i$ cannot remove a correlation with something it does not contain. The within-pair estimator is not exposed to this. Because both pitches of a pair are shown to the same respondent, the trait $\xi_i$ takes the same value in both potential outcomes, so the within-pair contrast holds it fixed and it never enters the estimand (Section~\ref{sec:withinpair}). Covariate adjustment cannot do the same, since it can condition only on the observed $X_i$ and not on $\xi_i$. Figure~\ref{fig:ex-unobs} shows the two responses: as the recommender sorts more strongly on the unobserved trait, the unconfoundedness-OLS estimate moves away from $\beta_c$, while the within-pair estimate stays at $\beta_c$.

\subsection{Negative Sample Comparisons}\label{sec:negex}

A common way to estimate feature effects in a recommender system is the \emph{negative-example} approach: treat the observed successes as positive examples, construct a set of negative examples to compare them against, and read the effect of a feature off the difference between positives and negatives. Because the positives are the successes, the approach conditions on the outcome from the outset; what it recovers then depends entirely on where the negative examples come from.

One construction uses the logged runner-up, the pitch that was offered to the respondent but not shown. Taking each funded, shown pitch as a positive and its non-shown pair-mate as the negative gives the \emph{funded-only comparison}. Because only funded interactions enter, it conditions on $Y_i=1$, an event realized after the display; conditioning on it opens a collider path between the display and the outcome, and the comparison is biased even when the display is randomized and there is no selection to correct.

A second construction keeps the unsuccessful interactions as well, and treats the non-shown pair-mate as the alternative the respondent chose against, the outside option in the language of demand estimation. Using both the funded and the unfunded interactions this way reproduces the within-pair transformed outcome of Section~\ref{sec:withinpair}, our preferred estimator, with one difference: it is applied to every offered pair, including the pairs with equal $v_c$. Retaining those pairs is exactly the case of Section~\ref{sec:ties}, where they dilute the estimand and, under the alternative weight, can reverse its sign.

A third construction does not use the runner-up at all; it draws negatives from the catalog at large and assigns them a zero outcome. Its probability limit, $\mathbb{E}[V_{i,c}\mid Y_i=1,\text{shown}]-\mathbb{E}[V_{i,c}\mid\text{catalog}]$, does not reduce to the within-pair moments: it compares the concreteness of pitches that were displayed and funded with that of catalog pitches that were never displayed and are assigned a zero outcome by construction, so the difference reflects which pitches the recommender chose to show, and the imputed zero, as much as any effect of concreteness on funding. It is not a causal contrast.

\subsection{Heterogeneity in the concreteness effect}\label{sec:het-example}

The within-pair estimator returns a single averaged effect; a natural next question is whether that effect varies across respondents. To study this, let the response also carry a quadratic term in concreteness,
\[
m(v,x,u)=\alpha_0+\beta_c v_c+\tfrac12\beta_{cc}v_c^2+\beta_s v_s+\eta_c x_c+\delta_{cc}v_c x_c+\psi u.
\]
Applying the within-pair estimator of Section~\ref{sec:withinpair} exactly as before, the per-unit contrast for a pair is $\beta_c+\beta_{cc}\bar V_{i,c}+\delta_{cc}X_{i,c}$ up to bundling, where $\bar V_{i,c}=\tfrac12(V_{i,c}^{h}(1)+V_{i,c}^{h}(-1))$ is the pair's average concreteness. Two terms make it vary, and they are different things. The interaction $\delta_{cc}X_{i,c}$ is genuine heterogeneity: the effect depends on the respondent's preference, and it does so whether or not the recommender sorts. The curvature term $\beta_{cc}\bar V_{i,c}$ is not heterogeneity across respondents; it says the effect depends on where on the concreteness scale the pair sits.

\begin{figure}[tbp]\centering
\includegraphics[width=\linewidth]{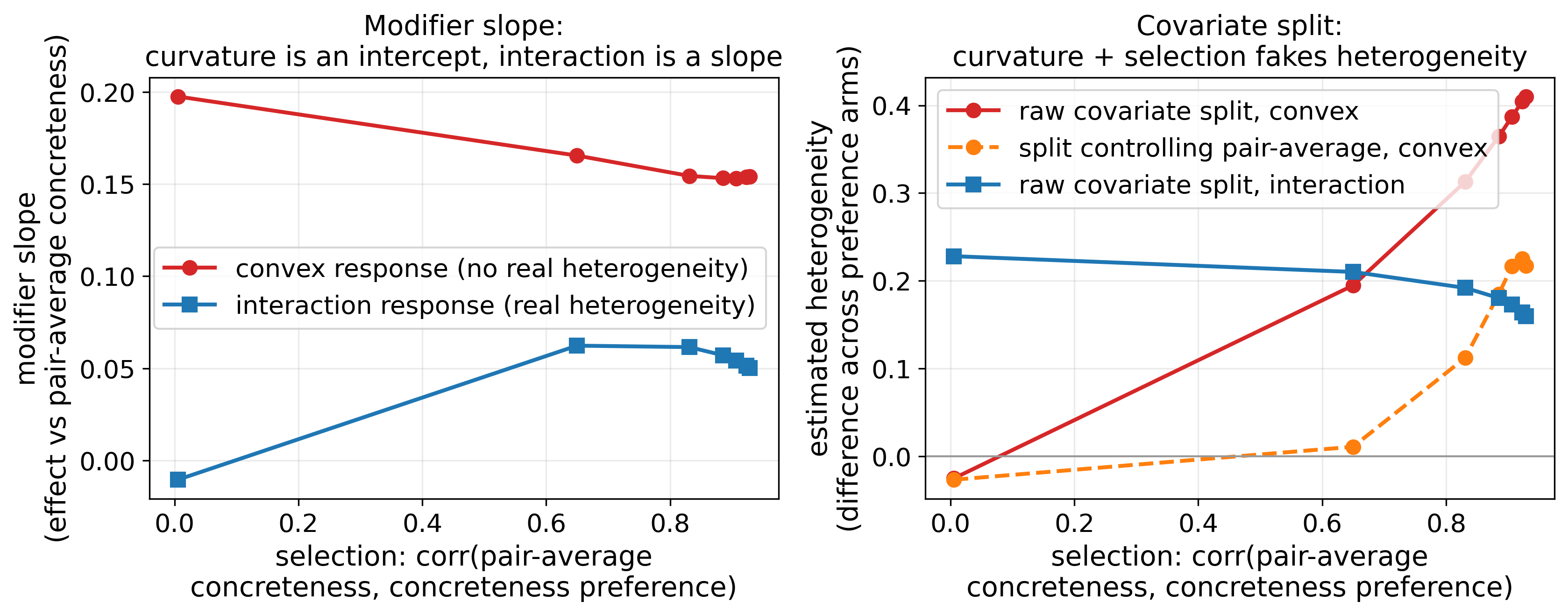}
\caption{Real versus spurious heterogeneity. Under a curved response with sorting, an uncontrolled split of the within-pair contrast across preference groups rises with selection although there is no true heterogeneity; under a genuine interaction the split is positive and unchanged by selection. Holding the pair's average concreteness fixed, or reading the random-assignment regime, recovers the genuine component.}
\label{fig:ex-het}
\end{figure}

The two are easily confused, and separating them is the point of Figure~\ref{fig:ex-het}. Suppose one estimates heterogeneity by relating the within-pair contrast to the respondent's preference $X_{i,c}$. Because the recommender offers more-concrete pairs to respondents with a higher preference for concreteness, $\bar V_{i,c}$ and $X_{i,c}$ move together, so a contrast that varies $X_{i,c}$ also varies $\bar V_{i,c}$ and picks up the curvature term as if it were respondent heterogeneity. To isolate the genuine component $\delta_{cc}$, hold the pair's average concreteness fixed by including $\bar V_{i,c}$ as a control; equivalently, read the relationship under random assignment, where $\bar V_{i,c}$ and $X_{i,c}$ are unrelated.

\subsection{Two features: joint identification and a placebo test}\label{sec:twofeat}

\begin{figure}[tbp]\centering
\includegraphics[width=0.7\linewidth]{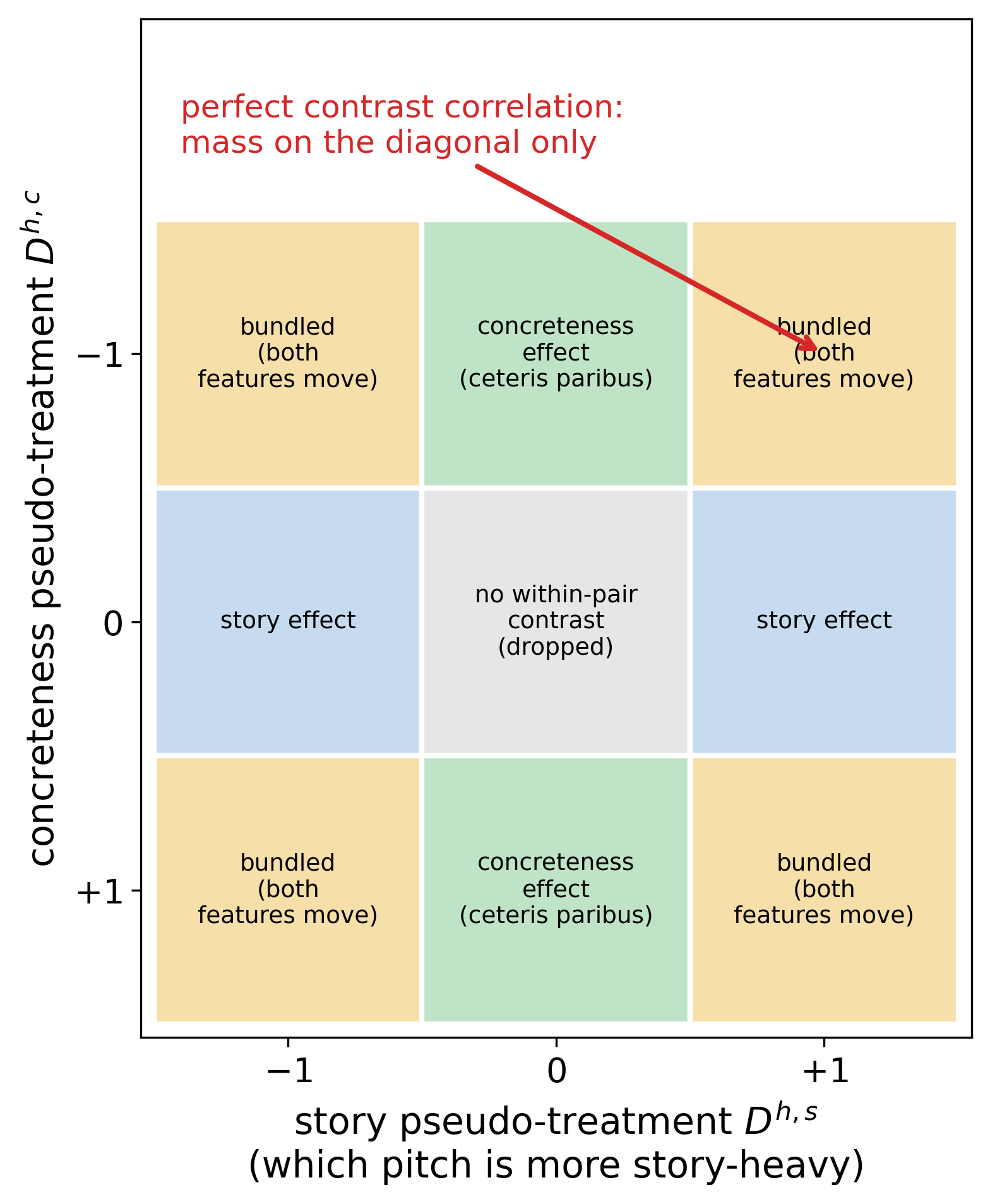}
\caption{Joint identification with two pseudo-treatments. Given variation in concreteness and story pseudo-treatments, a within-pair concreteness contrast identifies the ceteris-paribus effect only where the story contrast is zero (concreteness differs, story tied); where both move together the contrast bundles the two features.}
\label{fig:ex-twotcf}
\end{figure}

\emph{Two treatment-characterizing functions.} With two features, one randomization yields two pseudo-treatments, one contrasting concreteness and one contrasting story. The identification result of Section~\ref{sec:multi-tcf} specializes here: the concreteness effect is identified holding story fixed only on the pairs that differ in concreteness but are equal in story, and where the recommender always moves the two features together, the ceteris-paribus concreteness effect is not separately identified without the story contrast. Figure~\ref{fig:ex-twotcf} maps the identified region as a function of of the two pseudo-treatments.

\begin{figure}[tbp]\centering
\includegraphics[width=\linewidth]{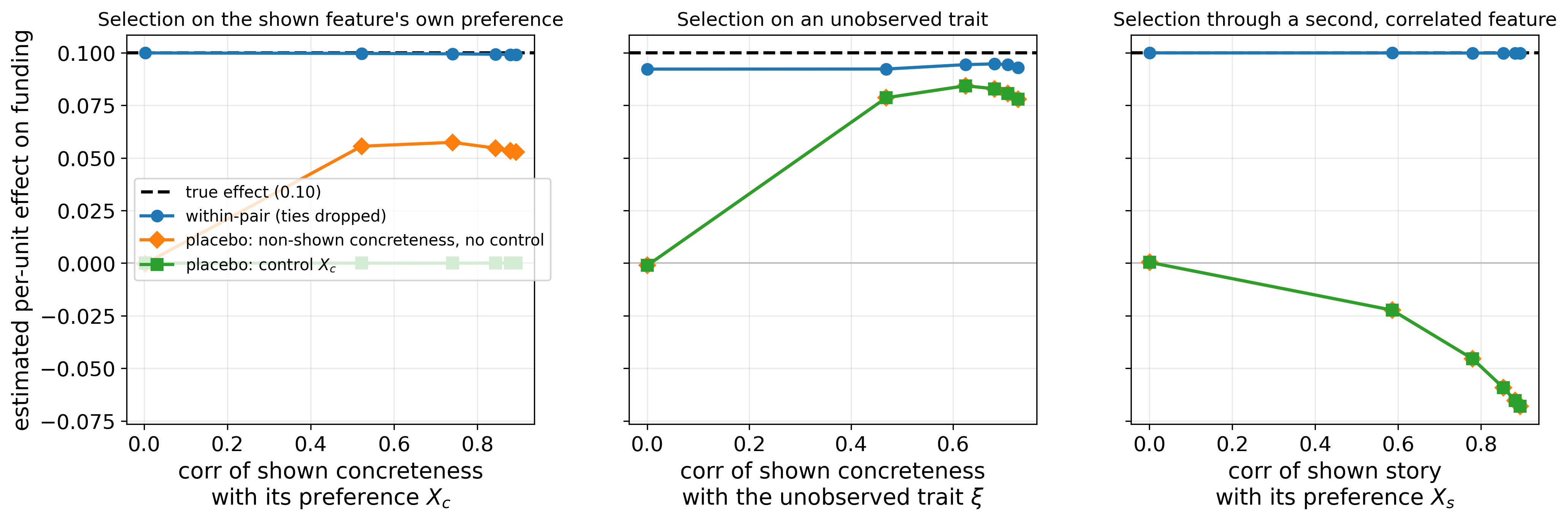}
\caption{A placebo on the non-shown pitch. The estimated effect of the non-shown pitch's concreteness, which is zero in truth, grows with selection. Controlling the observed covariate removes it when the selection runs through that covariate (left), but not when it runs through an unobserved trait (middle) or a second, correlated feature (right).}
\label{fig:ex-placebo}
\end{figure}

\emph{A placebo test.} The non-shown pitch cannot have a causal effect on funding, so a regression that finds a non-zero effect is detecting selection, not an effect. Regress $Y_i$ on a constant, the shown concreteness $V_{i,c}$, and the non-shown concreteness $V_{i,c}^N$, and let $\pi^N$ be the coefficient on $V_{i,c}^N$. Because $V_{i,c}^N$ does not enter the response $m$, the only way it can predict $Y_i$ is through its correlation, once the shown concreteness is partialled out, with the terms that do enter. Writing $V_{i,c}^{N\perp}$ for the residual of $V_{i,c}^N$ on $(1,V_{i,c})$ and $\varphi_z=\operatorname{Cov}(V_{i,c}^{N\perp},z)/\operatorname{Var}(V_{i,c}^{N\perp})$ for the partial regression coefficient of a funding driver $z$ on it,
\[
\pi^N=\eta_c\,\varphi_{X_{c}}+\psi\,\varphi_{U}+\beta_s\,\varphi_{V_{s}},
\]
one term for each determinant of funding with which the non-shown pitch is correlated (a further term in $\delta_{cc}$ appears when the effect is heterogeneous). Under a randomized display with no selection, $V_{i,c}^N$ is unrelated to the respondent's preference $X_{i,c}$ and trait $U_i$ and to the shown pitch's story $V_{i,s}$, so each $\varphi$ is zero and $\pi^N=0$; a nonzero $\pi^N$ therefore signals selection. The three terms are the three routes selection can take: through the observed preference ($\eta_c\varphi_{X_c}$), through an unobserved trait ($\psi\varphi_{U}$), and through a correlated second feature ($\beta_s\varphi_{V_s}$). Adding $X_i$ to the regression removes the first term but leaves the other two. Figure~\ref{fig:ex-placebo} shows $\pi^N$ growing with the strength of selection, and shows that controlling $X_i$ removes it only when the selection runs through $X_i$ rather than through an unobserved trait or a second feature.

\section{Evaluating Large Language Models}

One natural way to compare the performance of Large Language Models (LLMs) is to present individuals with responses from two or more LLMs to a prompt and ask them to rank the responses. When the responses are long, and especially when they involve repeated interactions between the individual and the LLM, this may be difficult to implement in practice. Here we show how the insights from the current paper can be exploited to evaluate LLMs indirectly.

Suppose one wishes to compare two LLMs, say A and B. An individual sends a prompt. Both LLMs receive the prompt and prepare a response. However, only one of the responses, selected at random, is forwarded to the individual. Now instead of asking the individual to rank two responses, the individual is asked to rate the response. To fit into the LCE framework developed in this paper, the data are now  the rating $Y_i$, features of the LLM response presented $V_i$,  features of the LLM response not presented $V^N_i$, and possibly individual and prompt characteristics $X_i$, in total the quadruple $(Y_i,V_i,V^N_i,X_i)$.

\section{Comparison with Unconfoundedness Analyses}\label{sec:unconf}

\subsection{The Importance of Dropping Units with $D_i^h=0$}

Given the data $(Y_i,V_i,V^N_i,X_i)$ an alternative strategy is just to compare units by values of the realized feature $V_i$. Suppose $V_i$ is binary, $V_i\in\{0,1\}$, and suppose we look at the difference in means estimator,
$\hat\tau^{\rm unc}=\overline{Y}_{V_i=1}-\overline{Y}_{V_i=0}$, the difference in average outcomes by the realized feature $V_i$.
With $h(v,v')=v-v'$, the definition of $D^h_i$ 
\[D_i^h\equiv 
\left\{
\begin{array}{rl}
1\hskip1cm & {\rm if}\ h(V_i,V_i^N)>0,\\
0\hskip1cm &  {\rm if}\ h(V_i,V_i^N)=0,\\
-1\hskip1cm &  {\rm if}\ h(V_i,V_i^N)<0,
\end{array}\right.\]
implies
\[D_i^h=
\left\{
\begin{array}{rl}
1\hskip1cm & {\rm if}\ V_i=1,V^N_i=0,\\
0\hskip1cm &  {\rm if}\ V_i=V_i^N,\\
-1\hskip1cm &  {\rm if}\ V_i=0,V_i^N=1.
\end{array}\right.\]
Now if we exclude the units with $D_i^h=0$, it implies that
\[ D^h_i=2V_i-1=1-2V^N_i,\]
and within the subpopulation with $D_i^h\neq 0$ $D_i^h$ and $V_i$ are perfectly correlated.
Thus:
\[ \hat \tau=\overline{Y}_{D^h_i=1}-\overline{Y}_{D^h_i=-1}=
\overline{Y}_{V_i=1,D_i^h\neq 0}-\overline{Y}_{V_i=0,D^h_i\neq 0}.
\]
Thus looking at the difference in means by $V_i$ is fine, as long as it is done within the subpopulation with $D^h_i\neq 0.$ In particular it means that for any unit-level covariates we must have in expectation
\[
\mathbb{E}\left[\overline{X}_{D^h_i=1}-\overline{X}_{D^h_i=-1}\right]=
\mathbb{E}\left[\overline{X}_{V_i=1,D_i^h\neq 0}-\overline{X}_{V_i=0,D^h_i\neq 0}\right]=0
\]

However, there is not in general a causal interpretation of $\overline{Y}_{V_i=1}-\overline{Y}_{V_i=0}$ without conditioning on $D_i^h=0$. Let us explore that difference in means. Expanding the unconditional difference in means will yield three terms: a causal contrast on the discordant subpopulation, scaled by a conditional probability, plus two bias terms, one arising from the imbalance of $V_i$ and one from the incomparability of the two tied subpopulations. The expansion is as follows.
\[\overline{Y}_{V_i=1}-\overline{Y}_{V_i=0}\]
\[
=\overline{Y}_{V_i=1,D_i^h\neq 0} \hat P(D_i^h\neq 0|V_i=1)
+\overline{Y}_{V_i=1,D_i^h= 0} \hat P(D^h_i=0|V_i=1)
\]\[\hskip2cm-\left(\overline{Y}_{V_i=0,D^h_i\neq 0}\hat P(D^h_i\neq 0|V_i=0)
+\overline{Y}_{V_i=0,D^h_i= 0}\hat P(D^h_i=0|V_i=0)
\right)
\]
\[
=\overline{Y}_{V_i=1,D_i^h\neq 0} \hat P(V^N_i=0|V_i=1)
+\overline{Y}_{V_i=1,D_i^h= 0} \hat P(V^N_i=1|V_i=1)
\]\[\hskip2cm -\left(\overline{Y}_{V_i=0,D^h_i\neq 0}\hat P(V^N_i=1|V_i=0)
+\overline{Y}_{V_i=0,D^h_i= 0}\hat P(V^N_i=0|V_i=0)
\right)
\]
\[
=\overline{Y}_{V_i=1,D_i^h\neq 0} \hat P(V^N_i=0|V_i=1)
+\overline{Y}_{V_i=1,D_i^h= 0} (1-\hat P(V^N_i=0|V_i=1)
\]\[\hskip2cm -\left(\overline{Y}_{V_i=0,D^h_i\neq 0}\hat P(V_i=0|V^N_i=1)\frac{\hat P(V^N_i=1)}{\hat P(V_i=0)}
+\overline{Y}_{V_i=0,D^h_i= 0}\left(
1-\hat P(V_i=0|V^N_i=1)\frac{\hat P(V^N_i=1)}{\hat P(V_i=0)}
\right)
\right).
\]
By symmetry $P(V_i=0)=P(V^N_i=0)$ and 
$P(V_i=0|V^N_i=1)=P(V^N_i=0|V_i=1)$, so this simplifies to
\[
=\overline{Y}_{V_i=1,D_i^h\neq 0} \hat P(V^N_i=0|V_i=1)
+\overline{Y}_{V_i=1,D_i^h= 0} (1-\hat P(V^N_i=0|V_i=1)
\]\[\hskip2cm -\left(\overline{Y}_{V_i=0,D^h_i\neq 0}\hat P(V^N_i=0|V_i=1)\frac{\hat P(V_i=1)}{\hat P(V_i=0)}
+\overline{Y}_{V_i=0,D^h_i= 0}\left(
1-\hat P(V^N_i=0|V_i=1)\frac{\hat P(V_i=1)}{\hat P(V_i=0)}
\right)
\right)
\]

\[
=\overline{Y}_{V_i=1,D_i^h\neq 0} \hat P(V^N_i=0|V_i=1)
-\overline{Y}_{V_i=0,D^h_i\neq 0}\hat P(V^N_i=0|V_i=1)\frac{\hat P(V_i=1)}{\hat P(V_i=0)}
\]\[
+\overline{Y}_{V_i=1,D_i^h= 0} (1-\hat P(V^N_i=0|V_i=1))
-\overline{Y}_{V_i=0,D^h_i= 0}\left(
1-\hat P(V^N_i=0|V_i=1)\frac{\hat P(V_i=1)}{\hat P(V_i=0)}
\right)
\]

\[
=\left(\overline{Y}_{V_i=1,D_i^h\neq 0}-\overline{Y}_{V_i=0,D^h_i\neq 0}\right) \hat P(V^N_i=0|V_i=1)
\]
\[+\overline{Y}_{V_i=0,D^h_i\neq 0}\hat P(V^N_i=0|V_i=1)\left(1-\frac{\hat P(V_i=1)}{\hat P(V_i=0)}\right)
\]\[
+\overline{Y}_{V_i=1,D_i^h= 0} (1-\hat P(V^N_i=0|V_i=1))
-\overline{Y}_{V_i=0,D^h_i= 0}\left(
1-\hat P(V^N_i=0|V_i=1)\frac{\hat P(V_i=1)}{\hat P(V_i=0)}
\right)
\]
The first of the three terms has a causal interpretation (though scaled by $ \hat P(V^N_i=0|V_i=1)$). The last two terms are bias terms. One comes from the fact that the distribution of $V_i$ is not balanced, and $\hat P(V_i=1)\neq \hat P(V_i=0). $
The second bias term (third term overall) comes from the fact that in the $V_i=1$ subpopulation we have units with $V_i=V^N_i=1$ and there is nothing in the model that makes these units comparable to other units, in particular to units with $V_i=V^N_i=0$. The difference between the size of the $V_i=V^N_i=1$ subpopulation versus the size of the $V_i=V^N_i=0$ 
subpopulation also changes the weights we put on the 
the subpopulations we do want to compare, units with $V_i=1$ and $V_i=0$ in the $D_i^h\neq 0$ subpopulation, leading to the first bias term.

\subsection{Unconfoundedness}

A common strategy is to use the data on the features of the treatment received, $V_i$, and analyze the data as if these characterize the treatment or causal variable. That is, the researcher assumes these features define potential outcomes $Y_i(v)$.  \citet{vanderweele2013causal} refer to this as the ``no-multiple-versionsof-the-treatment. Given these potential outcomes $Y_i(v)$ it is then often assumed that assignment to these features is unconfounded given the individual characteristics:
\[ V_i\ \indep\ Y_i(v)\ \Bigl|\ X_i.\]
This validates  estimating the average causal effect of changing the treatment from $v$ to $v'
$ as
\[ \tau(v,v')=\mathbb{E}
\left[\mathbb{E}\left[\left.Y_i\right|V_i=v',X_i\right]\right]-
\mathbb{E}
\left[\mathbb{E}\left[\left.Y_i\right|V_i=v,X_i\right]\right].
\]
The question we address here is how this approach relates to the approach developed in the current paper.

The first point is that this unconfoundedness approach does not explicitly use data on (features of) the treatment not received, $V^N_i$ and $A_i^N$. It only uses data on the triple $(X_i,V_i,Y_i)$. Thus, our proposed approach uses additional data, and it is therefore not surprising it may be able to relax assumptions commonly made, or answer additional questions.

Second, this approach implicitly assumes that the features fully characterize the treatment, so that essentially there is no distinction between the treatment $A_i$ and the features $V_i$, or $V_i=A_i.$
We can weaken this slightly by making a surrogacy type assumption, {\it e.g.,} \cite{athey2025surrogate}, requiring independence of the treatments and the outcome given features of the treatment and the covariates.

\section{Conclusion}\label{sec:conclusion}

This paper makes three contributions. First, we define Logged Counterfactual Exposures, data that augments standard interaction logs with content that could have been shown but was not, together with the relative exposure probability of the pair. Second, we introduce estimands, Conditional Treatment-averaged Causal Effects, that average unit-level contrasts over both users and the pairs of items a stochastic algorithm might have exposed to them, and we show these estimands are identified from LCE data even when unobserved confounders drive both user preferences and the composition of the pair of potential exposures. Our estimator requires neither a model of the assignment mechanism nor observation of the full set of content features. Third, we show that LCE data is practically obtainable at scale: from the token-level log-probabilities of generative models, from replaying stochastic algorithms at logged configurations, and from the candidate sets and experimentation infrastructure of production recommendation systems. Thus, data that may be collected in the regular course of business, or that may be augmented without obtaining additional interactions with users, can be used to estimate causal effects from observational data. 

The estimand has a subtle interpretation. A C-TACE is the average effect of exposing users to higher-feature content \emph{as the deployed algorithm produces it}, inclusive of the content dimensions that are correlated with the feature in that algorithm's output. For a platform deciding whether to steer its algorithm toward the feature, this is the policy-relevant quantity. The linear example of Section \ref{sec:linear-example} makes precise how it relates to the isolated per-feature effect and when the two coincide. Because the intrinsically stochastic nature of modern AI systems generates micro experiments in the course of ordinary use, the data our estimands require can be easy to obtain. Analyses of this kind can thus serve as a useful complement to platform experimentation.

\bibliography{\bib}

\appendix

\section{Derivations}\label{sec:example-derivations}

Throughout, expectations are over the population, $\bar h=\mathbb{E}[\Delta_i\mid\Delta_i>0]$, and $\theta_m=(\Sigma_V)_{cm}/(\Sigma_V)_{cc}$ is the coefficient of feature $m$'s within-pair gap on the concreteness gap, so $B_0=\beta_c+\sum_{m\ne c}\theta_m\beta_m$.

\emph{Within-pair.} Fix the offered pair and the respondent's $(X_i,U_i)$. When the more-concrete pitch is displayed ($D_i^h=+1$) the transformed outcome is $\tilde Y_i=2Y_i$ and $\mathbb{E}[Y_i\mid\cdot]=m(V_i^{h+},X_i,U_i)$; when the less-concrete pitch is displayed ($D_i^h=-1$) it is $\tilde Y_i=-2Y_i$ and $\mathbb{E}[Y_i\mid\cdot]=m(V_i^{h-},X_i,U_i)$. Since the display shows the more-concrete pitch with probability $P$,
\[
\mathbb{E}[\tilde Y_i\mid\text{pair},X_i,U_i]
=2P\,m(V_i^{h+},X_i,U_i)-2(1-P)\,m(V_i^{h-},X_i,U_i),
\]
which at $P=\tfrac12$ equals $m(V_i^{h+},X_i,U_i)-m(V_i^{h-},X_i,U_i)$. Substituting the linear response, the intercept $\alpha_0$, the covariate term $\eta_c X_{i,c}$, and the unobserved-trait term $\psi U_i$ take the same value at $V_i^{h+}$ and $V_i^{h-}$ (they do not depend on the pitch), so they leave the difference unchanged:
\[
m(V_i^{h+},X_i,U_i)-m(V_i^{h-},X_i,U_i)=\beta_c\,\Delta_i+\beta_s\big(V_{i,s}^{h}(1)-V_{i,s}^{h}(-1)\big)+\delta_{cc}\,\Delta_i\,X_{i,c}.
\]
Taking the expectation over the discordant pairs and dividing by $\bar h$ gives the per-unit estimand $B_0$: the $\beta_s$ term contributes $\theta_s\beta_s$ because the story gap moves with the concreteness gap in the bank, and the $\delta_{cc}$ term averages to a bundled component as well.

\emph{Naive OLS.} The regression of $Y_i$ on a constant and $V_{i,c}$ over the shown sample has slope $\operatorname{Cov}(V_{i,c},Y_i)/\operatorname{Var}(V_{i,c})$ in the limit. Replacing $Y_i$ by its conditional mean $m(V_i,X_i,U_i)$ and using the linearity of covariance,
\[
\operatorname{Cov}(V_{i,c},Y_i)=\beta_c\operatorname{Var}(V_{i,c})+\sum_{m\ne c}\beta_m\operatorname{Cov}(V_{i,c},V_{i,m})+\eta_c\operatorname{Cov}(V_{i,c},X_{i,c})+\psi\operatorname{Cov}(V_{i,c},U_i).
\]
Dividing by $\operatorname{Var}(V_{i,c})$ turns the first two groups of terms into $B_0=\beta_c+\sum_{m\ne c}\theta_m\beta_m$ and the last two into $\eta_c b+\psi\,\operatorname{Cov}(V_{i,c},U_i)/\operatorname{Var}(V_{i,c})$, where $b=\operatorname{Cov}(V_{i,c},X_{i,c})/\operatorname{Var}(V_{i,c})$ is the regression coefficient of the respondent's preference on the shown concreteness. Under selection on observables only, the last term is zero and the limit is $B_0+\eta_c b$.

\emph{Unconfoundedness OLS.} Now regress $Y_i$ on $V_{i,c}$ and the covariate vector $X_i$. By the Frisch--Waugh--Lovell theorem the coefficient on $V_{i,c}$ equals the univariate slope of $Y_i$ on $V_{i,c}^{\perp}$, the residual from the linear projection of $V_{i,c}$ on $(1,X_i)$, so the limit is $\operatorname{Cov}(V_{i,c}^{\perp},Y_i)/\operatorname{Var}(V_{i,c}^{\perp})$. Because $V_{i,c}^{\perp}$ is orthogonal to $X_i$ by construction, the $\eta_c X_{i,c}$ term of $m$ contributes nothing; the bundled term $B_0$ is unchanged, as it comes from the shown pitch's own correlated features. What remains is the unobserved-trait term:
\[
\operatorname{plim}\hat\beta^{\text{unc}}=B_0+\psi\,\frac{\operatorname{Cov}(V_{i,c}^{\perp},U_i)}{\operatorname{Var}(V_{i,c}^{\perp})},
\]
which is zero exactly when $U_i$ is uncorrelated with $V_{i,c}$ given $X_i$, that is, when the recommender did not sort on anything the analyst fails to observe.

\emph{Negative examples with the runner-up.} Center the shown concreteness at the pair mean, $c_i=V_{i,c}-\bar V_{i,c}=\tfrac12(V_{i,c}-V_{i,c}^N)$, and regress $Y_i$ on $c_i$. Conditional on the pair the display is a fair draw, so $c_i$ takes the values $\pm\tfrac12\Delta_i$ with equal probability; hence $\mathbb{E}[c_i\mid\text{pair}]=0$, $\operatorname{Var}(c_i\mid\text{pair})=\tfrac14\Delta_i^2$, and $\mathbb{E}[c_iY_i\mid\text{pair}]=\tfrac14\Delta_i\big(m(V_i^{h+},X_i,U_i)-m(V_i^{h-},X_i,U_i)\big)$. Taking expectations over pairs, the $\tfrac14\Delta_i^2$ factors cancel between numerator and denominator and $\operatorname{Cov}(c_i,Y_i)/\operatorname{Var}(c_i)=B_0$. Regressing on the raw difference $V_{i,c}-V_{i,c}^N=2c_i$ scales the slope by $\tfrac12$, returning $\tfrac12 B_0$. The funded-only version restricts to $Y_i=1$ and contrasts the shown and non-shown concreteness among the funded, with limit $\big(\operatorname{Cov}(V_{i,c},Y_i)-\operatorname{Cov}(V_{i,c}^N,Y_i)\big)/\mathbb{E}[Y_i]$; because it conditions on the outcome, it is nonzero even under a randomized display ($b=0$), a collider.

\emph{Ties kept.} A pair with equal $v_c$ has $D_i^h=0$ and $\tilde Y_i=0$, so it adds nothing to the numerator of the within-pair average. Under the natural weight the mean of $\tilde Y_i$ over all pairs is the mean over the discordant pairs times their share, $1-\pi_0$; dividing by $\bar h$ gives $(1-\pi_0)\tau_h$, the discordant-pair effect scaled down by the non-tie fraction. The alternative weight instead assigns each tied pair the weight $-1/(1-P)$ on its outcome $Y_i$, as though its less-concrete member had been displayed; averaging, this adds $-\tfrac{\pi_0}{1-P}\,\mathbb{E}[Y_i\mid\text{tie}]$ to the limit, a term fixed by the funding rate on tied pairs and unrelated to concreteness.

\end{document}